\documentclass[twocolumn,showpacs,preprintnumbers,superscriptaddress,
	     amsmath,amssymb,prd]{revtex4}

\usepackage[dvips]{color,graphicx}
\usepackage{dcolumn}
\usepackage{bm}
  \makeatletter
  \def\mathcomposite{%
     \@ifstar
        {\def\@mathcomposite@option{%
            \baselineskip\z@skip\lineskiplimit-\maxdimen}%
         \@mathcomposite}%
        {\let\@mathcomposite@option\offinterlineskip
         \@mathcomposite}}
  \def\@mathcomposite{%
     \@ifnextchar[\@@mathcomposite{\@@mathcomposite[0]}}
  \def\@@mathcomposite[#1]#2#3#4{%
     #2{\mathchoice
        {\@mathcomposite@{#1}{#3}{#4}\displaystyle{1}}%
        {\@mathcomposite@{#1}{#3}{#4}\textstyle{1}}%
        {\@mathcomposite@{#1}{#3}{#4}%
         \scriptstyle\defaultscriptratio}%
        {\@mathcomposite@{#1}{#3}{#4}%
         \scriptscriptstyle\defaultscriptscriptratio}}}
  \def\@mathcomposite@#1#2#3#4#5{%
     \vcenter{\m@th\@mathcomposite@option
        \dimen@\f@size\p@\dimen@#1\dimen@\dimen@#5\dimen@
        \divide\dimen@ 18
        \edef\@mathcomposite@skipamount{\the\dimen@}%
        \ialign{\hfil$#4##$\hfil\cr
           #2\crcr
           \noalign{\vskip\@mathcomposite@skipamount}%
           #3\crcr}}}
  \makeatother
\renewcommand{\agt}{\mathcomposite{\mathrel}{>}{\sim}}
\renewcommand{\alt}{\mathcomposite{\mathrel}{<}{\sim}}

\def\bfm#1{\mbox{\boldmath $#1$}}
\def\bfsm#1{\mathstrut\mbox{\scriptsize{\boldmath $#1$}}\mathstrut}

\newcommand{\al}{\alpha}
\newcommand{\be}{\beta}
\newcommand{\ga}{\gamma}
\newcommand{\Ga}{\Gamma}
\newcommand{\de}{\delta}
\newcommand{\De}{\Delta}
\newcommand{\ep}{\varepsilon}
\newcommand{\eps}{\epsilon}
\newcommand{\ze}{\zeta}
\newcommand{\ka}{\kappa}
\newcommand{\la}{\lambda}
\newcommand{\Lam}{\Lambda}

\newcommand{\si}{\sigma}

\renewcommand{\th}{\theta}  

\newcommand{\om}{\omega}
\newcommand{\Om}{\Omega}
\newcommand{\p}{\partial}
\newcommand{\lt}{\langle} 
\newcommand{\gt}{\rangle}  
\newcommand{\txt}{\textstyle}
\newcommand{\dsp}{\displaystyle}
\newcommand{\scr}{\scriptstyle}
\newcommand{\Eqn}[1]{Eq.~(\ref{#1})}  
\newcommand{\beq}{\begin{equation}}
\newcommand{\eeq}{\end{equation}}
\newcommand{\ba}{\begin{array}}
\newcommand{\ea}{\end{array}}
\newcommand{\bea}{\begin{eqnarray}}
\newcommand{\eea}{\end{eqnarray}}
\newcommand{\bal}{\begin{align}}  
\newcommand{\eal}{\end{align}}
\newcommand{\bi}{\begin{itemize}}  
\newcommand{\ei}{\end{itemize}}
\newcommand{\ben}{\begin{enumerate}}  
\newcommand{\een}{\end{enumerate}}

\newcommand{\half} {{\txt \frac{1}{2}}}
\newcommand{\third}{{\txt \frac{1}{3}}}

\newcommand{\twothirds}{{\txt \frac{2}{3}}}

\newcommand\hide[1]{}

\newcommand{\tr}{\mbox{tr}}

\newcommand{\ie}{{i.e.}}

\newcommand{\nn}{\nonumber \\}

\newcommand{\feyn}[1]{
  \setbox0=\hbox{\ensuremath{#1}}
  \hbox to\wd0{\hbox to0pt{\hbox to\wd0{\hss/\hss}\hss}\box0}}

\newcommand{\MeV}{\,{\rm MeV}} 
\newcommand{\GeV}{\,{\rm GeV}} 
\newcommand{\diag}{{\rm diag.}}
\newcommand{\Qtilde}{{\tilde Q}}
\newcommand{\mue}{\mu_{e}}
\newcommand{\ms}{m_s}
\newcommand{\Ms}{M_s}
\newcommand{\btem}{\bibitem}

\begin{document}


\title{Extensive study of phase diagram for charge neutral homogeneous
quark matter affected by dynamical chiral condensation\\
-- {\it unified picture for thermal unpairing transitions from weak to
        strong coupling}  --}
\author{Hiroaki Abuki}
\email[E-mail:~]{abuki@yukawa.kyoto-u.ac.jp}
\affiliation{Yukawa Institute for Theoretical Physics, Kyoto University, 
Kyoto 606-8502, Japan}
\author{Teiji Kunihiro}
\email[E-mail:~]{kunihiro@yukawa.kyoto-u.ac.jp}
\affiliation{Yukawa Institute for Theoretical Physics, Kyoto University, 
Kyoto 606-8502, Japan}

\date{\today}

\begin{abstract}
We study the phase structures of charge neutral quark matter under
 the $\be$-equilibrium for a wide range of the quark-quark coupling
 strength within a four-Fermion model.
A comprehensive and unified picture for the phase transitions
 from weak to strong coupling is presented.
We first develop a technique to deal with the gap equation and
 neutrality constraints without recourse to numerical derivatives,
and show that the off-diagonal color densities automatically
 vanish with the standard assumption for the diquark condensates.
The following are shown by the numerical analyses:
(i) The thermally-robustest pairing phase
 is the two-flavor pairing (2SC) in any coupling case, while the
 second one for relatively low density is the
 up-quark pairing (uSC) phase or the color-flavor locked (CFL) phase
 depending on the coupling strength and the value of strange quark mass.
(ii) If the diquark coupling strength is large enough, the phase diagram
 is much simplified and is free from the instability problems associated
 with imaginary Meissner masses in the gapless phases.
(iii) The interplay between the chiral and diquark dynamics may bring
 a non-trivial first order transition even in the pairing phases at high
 density.
We confirm (i) also by using the Ginzburg-Landau analysis expanding
 the pair susceptibilities up to quartic order in the strange quark
 mass.
We obtain the analytic expression for the doubly critical density where
 the two lines for the second order phase
 transitions merge, and below which the down-quark pairing (dSC) phase
 is taken over by the uSC phase.
Also we study how the phase transitions from fully gapped states to
 partially ungapped states are smeared at finite temperature by
 introducing the order parameters for these transitions.
\end{abstract}

\pacs{12.38.-t, 25.75.Nq}
\maketitle

\section{Introduction}

Surprisingly rich phase structure of quark matter is being revealed by
extensive theoretical studies
\cite{Bailin:1983bm,Iwasaki:1994ij,reviews}.
On the basis of the asymptotic-free nature of QCD and the attraction
between quarks due to a gluon exchange, it is now believed that the ground
state of the sufficiently cold and extremely dense matter is in
the color-flavor locked (CFL) phase where all the quark species equally
participate in pairing \cite{Alford:1998mk}. 
In contrast, it remains still controversial what phase sets in next to
the CFL as the density is decreased and/or the temperature is raised. 

At zero temperature ($T=0$), the following two ingredients play crucial
roles for determining the second densest phase in QCD;
(i) the strange quark mass
\cite{Alford:1999pa,Schafer:1999pb,Abuki:2003ut}, and 
(ii) the charge neutrality constraints as well as the
$\be$-equilibrium condition
\cite{Iida:2000ha,Rajagopal:2000ff,Alford:2002kj}. 
The former tends to bring a Fermi-momentum mismatch between the
light and strange quarks, while the latter in the paired phase tends to
match the Fermi-momenta by tuning the charge chemical potentials. 
Some theoretical studies suggest that, when the baryon density is
decreased, the effect (ii) in the CFL phase cancels the effect (i)
down to some critical density, at which the CFL phase turns into 
the gapless CFL (gCFL) phase \cite{Alford:2003fq}. 
In the realistic quark matter in compact stars, the effect of (ii) gives
so strong constraint because of a long range nature of gauge interactions
that such an exotic phase may exist stably. 
The gCFL phase has been extensively studied and claimed to be the most
promising candidate of next phase down in density at vanishing
temperature \cite{Alford:2003fq} and is also known to lead
an interesting astrophysical consequence on the cooling history of 
an aged compact star \cite{Alford:2004zr}. 
It should be, however, noted here that the gapless phases are usually
accompanied by some transverse gluonic modes with imaginary Meissner
mass at low temperature \cite{Unstable,Fukushima:2005cm}, indicating
the existence of more stable exotic states
\cite{Reddy:2004my,Huang:2005pv,Casalbuoni:2005zp,%
Giannakis:2005sa,Hong:2005jv,Gorbar:2005rx,Schafer:2005ym}.
Although the resolution of the instability or finding the
possible new stable state is now one of the central issues in 
this field,
we won't deal with this problem; nevertheless we will give some
suggestions to possible resolution on the basis of the results
obtained in this work.

The neutrality constraints are also known to bring about an interesting
phase even at finite temperature $(T\ne0)$.
In fact, a Ginzburg-Landau analysis shows  that the $d$ quark pairing
phase with the $u$-$d$ and $d$-$s$ pairings, denoted by ``dSC'',
appears as the second phase when the 2SC (CFL) phase is cooled (heated)
\cite{Iida:2003cc}.
The existence of such a kind of intermediate phase has also been
confirmed using the NJL model \cite{Fukushima:2004zq}.
Also the gapless version of the dSC with an additional gapless 
mode is also examined for finite temperature \cite{Iida:2003cc} and for
zero temperature \cite{Abuki:2004zk}; the gapless dSC is shown to be
always a metastable at zero temperature, while it may exist stably
for nonzero temperature.
It should be noted, however, that the gapless phase for nonzero
temperature cannot be thermodynamically distinguished from the fully
gapped one; in fact, it has no distinct phase boundary with the gapped
one, and thus is of less physical interest than that at zero
temperature.

In the most of the literature concerning the pairing phases of the quark 
matter \cite{Alford:2003fq,Fukushima:2004zq,Ruster:2004eg}, the strange
quark mass is treated as a parameter just like an external
magnetic field applied to a metallic superconductor \cite{Sarma}. 
In QCD, however, the strong attraction exists in the scaler
quark-antiquark channel which leads to a non-perturbative phenomenon
called the dynamical chiral condensation in the low density regime. 
It is interesting to investigate how the incorporation of the dynamical
formation of the chiral condensates affects color superconducting phases
\cite{abuki00,Kitazawa:2002bc}.
The competition between the chiral and diquark condensations under the
charge neutrality constraints was first investigated in a four-Fermion
model \cite{Abuki:2004zk}; it
was shown that the second densest phase next to the CFL strongly depends
on the quark-quark (diquark) coupling strength, and in particular, the
gapless CFL phase may be washed out from the phase diagram except for an
extremely weak coupling case.  
Their investigation is extended to finite temperatures
\cite{Ruster:2005jc,Blaschke:2005uj} and recently also applied to the
system with finite neutrino density \cite{Ruster:2005ib}.
Some of the results obtained in \cite{Ruster:2005jc,Blaschke:2005uj},
however, seemingly contradict with the previous claim by the authors of
\cite{Iida:2003cc,Fukushima:2004zq} on the appearance of the dSC phase.
This is one of puzzles indicating that a further investigation is 
needed for a systematic understanding of the phases of quark 
matter and transitions among them.

A main aim of this paper is to explore the phase diagram for a wide
range of diquark coupling systematically 
and thereby give a unified picture for the phase
transitions from weak to strong coupling.
We clarify the detailed mechanism and features of thermal phase
transitions, and make clear the relations among the previous studies
\cite{Iida:2003cc,Fukushima:2004zq,Ruster:2005jc,Blaschke:2005uj}.
In particular, we shall study the the following. 
(i) How the $(\mu,T)$-domain accommodating the pairing grows
and that for the gapless phases vanishes as the coupling in the scaler
diquark channel is increased.
(ii) How we can draw a unified picture for the thermal unpairing 
phase transitions from weak to strong coupling;
for this purpose, we extend the previous study \cite{Iida:2003cc}
by making re-analysis of Ginzburg-Landau expansion focusing on
the effects of the terms quartic in the strange quark mass. 
(iii) How and why non-trivial first order phase transitions could
be caused by the competition between the pairing and chiral dynamics in
the strange quark sector.
(iv) How the transition from the fully gapped phase to the partially
ungapped (gapless) one is smeared at finite temperature by
introducing a definite order parameter. 

The paper is organized as follows.
In Sec.~\ref{Formulation}, we introduce the model and derive the gap
equation under the neutrality constraints.
Some technical developments are presented, by which the numerical
derivatives can be replaced by simple algebraic equations.
In Sec.~\ref{Numericalresults}, we discuss the points listed above
based on the numerical results.
Also the Ginzburg-Landau analysis is performed and some interesting
aspects of the thermal phase transitions are given at the 
end of this section.
The summary and outlook are provided in Sec.~\ref{Summary}.
In Appendix \ref{GLapproach}, we derive the Ginzburg-Landau potential
expanded up to quartic order in the strange quark mass.
In Appendix \ref{Offdiagonal}, we show that the off-diagonal color
densities automatically vanish under the standard ansatz for
the diquark condensates.

\vspace*{7ex}
\section{Formulation}\label{Formulation}
In this section, we introduce our model and formulate the mean field
approximation under the kinetic constraints.
In Sec.~\ref{hamil}, we present some useful algebraic techniques
to solve the gap equation and constraints without recourse to numerical
derivatives.

\subsection{Model}\label{ourmodel}
We start with the following Lagrangian as in \cite{Abuki:2004zk}
\bea
 {\mathcal L}&=&\bar{q}(i\feyn{\p} - \bfm{m}_0 + \bfm{\mu}\ga_0)q%
     + \frac{G_d}{16}\sum_{\eta=1}^{3}%
       \big[(\bar{q}P_\eta^t\bar{q})(^tq\bar{P}_\eta q)\big]\nn
 & & + \frac{G_s}{8N_c}%
     \big[(\bar{q}\bfm{\la}_Fq)^2+ (\bar{q}i\ga_5\bfm{\la}_F q)^2\big].\label{eq:lag}
\eea
Here, $\bfm{\la}_F=\{\sqrt{2/3}\bfm{1},\vec{\la}_F\}$ are the unit
matrix and the Gell-Mann matrices in the flavor space. 
$P_\eta$ is defined as \cite{Alford:2003fq}
\beq
  (P_\eta)^{ab}_{ij}=i\ga_5C\eps^{\eta ab}\eps_{\eta ij}%
  \quad\mbox{no sum over index $\eta$}\,
\label{assumption}
\eeq
and $\bar{P}_\eta=\ga_0 (P_\eta)^\dagger\ga_0$. 
$a,b,\cdots$ and $i,j,\cdots$ represent the color and flavor indices,
 respectively. 
The second term in \Eqn{eq:lag} simulates the attractive interaction in
  the color anti-triplet, flavor anti-triplet and $J^P=0^+$
  channel in QCD. 
$\bfm{m}_0=\diag\{m_u,m_d,\ms\}$ is the current-quark mass matrix.
In order to impose the color and electric neutralities, we have introduced
in \Eqn{eq:lag} the chemical potential matrix $\bfm{\mu}$ in the
color-flavor space as
\beq
\ba{rcl}
 \bfm{\mu}_{ij}^{ab}&=& \dsp \mu - \mue Q_{ij} + \mu_3 T_3^{ab} + \mu_8
                 T_8^{ab}\\[1ex]
         & &\dsp + \mu_1 T_1^{ab} + \mu_2 T_2^{ab} + \mu_4 T_4^{ab}\\[1ex] 
	 & &+ \mu_5 T_5^{ab} + \mu_6 T_6^{ab} + \mu_7 T_7^{ab}.
\ea
\label{chemicalpot}
\eeq
$Q=\diag(2/3,-1/3,-1/3)$ counts electric charge of each quark
species. 
$T_3=\diag(1/2,-1/2,0)$ and $T_8=\diag(1/3,1/3,-2/3)$ are
the diagonal charges for quarks in the fundamental representation of the
color SU(3).
We have included the second and third lines in \Eqn{chemicalpot} which
stand for the chemical potentials for off-diagonal color charges:
It has been recently claimed that 
these chemical potentials should be included for
the complete color neutralities
\cite{Buballa:2005bv}.
We can prove, however, that if the diquark condensates have
the color-flavor structure as given by \Eqn{assumption}, 
then the off-diagonal color densities must automatically vanish for
$\mu_1=\mu_2=\mu_4=\mu_5=\mu_6=\mu_7=0$; 
see Appendix \ref{Offdiagonal} for the detail of the proof.
Thus, we can safely adopt the usual diagonal ansatz for the chemical
potential matrix.
The explicit forms of the diagonal elements ($\mu_{ai}=\bfm{\mu}^{aa}_{ii}$)
are as follows.
\beq
\ba{rcl}
 \mu_{ru}&=& \mu-\twothirds\mue+\half\mu_3+\third\mu_8,\\[1ex]
 \mu_{gd}&=& \mu+\third\mue-\half\mu_3+\third\mu_8,\\[1ex]
 \mu_{bs}&=& 
\mu+\third\mue-\twothirds\mu_8,\\[3ex]
\mu_{rd}&=& \mu+\third\mue+\half\mu_3+\third\mu_8,\\[1ex]
 \mu_{gu}&=& \mu-\twothirds\mue-\half\mu_3+\third\mu_8,\\[3ex]
\mu_{rs}&=& 
\mu+\third\mue+\half\mu_3+\third\mu_8,\\[1ex]
 \mu_{bu}&=& \mu-\twothirds\mue-\twothirds\mu_8,\\[3ex]
\mu_{gs}&=& 
\mu+\third\mue-\half\mu_3+\third\mu_8,\\[1ex]
 \mu_{bd}&=& \mu+\third\mue-\twothirds\mu_8\ . \\[1ex]
\ea
\label{mudefinitions}
\eeq

We treat the diquark coupling constant $G_d$ as a simple parameter,
  although the perturbative one-gluon exchange vertex ${\cal L}_{\rm
  int}=-(g^2/2)\bar{q}\ga_\mu(\la_a/2)q\bar{q}\ga^\mu(\la_a/2)q$, 
  which is valid at extremely high density, tells us that $G_d/G_s=1/2$
  with $N_c=3$ \cite{Steiner:2002gx,Buballa:2001gj,Buballa:2003qv}.
As a measure of coupling constant $G_d$, we shall mainly use the
  gap energy ($\De_0$) in the pure CFL phase at $\mu=500\MeV$ and $T=0$
  in the chiral SU(3) limit, as in
  \cite{Alford:2003fq,Fukushima:2004zq,Abuki:2004zk}.

We evaluate the thermodynamic potential in the mean-field approximation;
\beq
\ba{rcl}
  \Om&=&\Om_{q} + \Om_e,\\[2ex]
  \Om_{q}&=&\dsp\frac{4}{G_d}\sum_{\eta=1}^{3}\De_\eta^2+%
  \frac{N_c}{G_s}\sum_{i=1}^{3}(M_i-m_i)^2 \\[2ex]
  &&\dsp-\frac{T}{2}\sum_{n}\int\frac{d\bfm{p}}{(2\pi)^3}\tr{\rm Log}%
    \left[S^{-1}(i\om_n,\bfm{p})\right],\\[1ex]
\ea
\label{det}
\eeq
where 
\bea
  \De_\eta&=&\frac{G_d}{8}\lt ^tqP_\eta q\gt,\\
  \bfm{M}&=&\left( \begin{array}{ccc}
  M_u& 0 & 0 \\
  0& M_d & 0 \\
  0& 0 & \Ms
  \end{array}\right)\nn
  &=&\bfm{m}_0-\frac{G_s}{2N_c}%
  \left( \begin{array}{ccc}
  \lt\bar{u}u\gt& 0 & 0  \\
  0 & \lt\bar{d}d\gt& 0 \\
  0 & 0 &\lt\bar{s}s\gt 
  \end{array}\right),\label{model}
\eea
are the gap parameter and constituent quark mass matrices.
$S$ denotes the $72\times 72$ Nambu-Gor'kov propagator defined by
\beq
   S^{-1}(i\om_n,p)=\left( \begin{array}{cc}
  \feyn{p}+{\bfm{\mu}}\ga_0-\bfm{M}& \sum_\eta
   P_\eta\De_\eta  \\
  \sum_\eta \bar{P}_\eta\De_\eta & 
  ^t\feyn{p}-{\bfm{\mu}}\ga_0+\bfm{M}
  \end{array}\right),\label{model2}
\eeq
   with $\feyn{p}=i\om_n\ga_0-\bfm{p}\cdot\bfm{\ga}$.
Finally, $\Om_e$ is the contribution from massless electrons
\bea
  \Om_e&=&-\frac{\mue^4}{12\pi^2}-\frac{\mue^2 T^2}{6}-\frac{7\pi^2 T^4}{180}.
\eea
The optimal values of the variational parameters $\De_\eta$, $M$ and $\Ms$
   must satisfy the stationary condition (the gap equations);
\beq
  \frac{\p\Om}{\p\De_\eta}=0,\,\,\frac{\p\Om}{\p M_{u,d}}=0%
  \,\,\mbox{and}\,\,\frac{\p\Om}{\p \Ms}=0.\label{eq:gapeq}
\eeq
Our task is to search the minimum of the effective potential by
   solving these  gap equations under the local electric and color
   charge neutrality conditions;
\beq
  \frac{\p\Om}{\p\mue}=0,\,\,\frac{\p\Om}{\p\mu_3}=0%
  \,\,\mbox{and}\,\,\frac{\p\Om}{\p\mu_8}=0.\label{eq:neut_con}
\eeq
For a later convenience, we define here the electron density by
\beq
  n_e=-\frac{\p\Om_e}{\p\mue}=\frac{\mue^3}{3\pi^2}+\frac{\mue T^2}{3}.
\eeq
The formulation made above is a straightforward extension of our
   previous work \cite{Abuki:2004zk} to the $(T\ne 0)$ case
   \cite{Ruster:2005jc,Blaschke:2005uj}.

\subsection{Representation of gap equation and kinetic constraints in
   terms of quasi-quark wave functions}\label{hamil}
In this section, we present some analytical way to deal with the gap
equation and neutrality constraints.
We shall show that the gap equation and neutrality constraints derived
above can be further simplified with the aid of the quasi-quark
wave functions (spinors).
By doing this, not only the physical meanings of these equations become
transparent, but also the numerical derivatives can be circumvented.
In particular, the latter has a practical advantage because the
computations of matrix elements are more favorable than the numerical
derivatives.

First, we introduce the Nambu-Gor'kov mean field Hamiltonian density
${\mathcal H}(\bfm{p};\bfm{\mu},\bfm{\De},\bfm{M})$ following
\cite{Alford:2003fq,Fukushima:2004zq},
\bea
  \Ga_0S^{-1}(i\om_n,\bfm{p})\bar{\Ga}_0=i\om_n{\bf 1}_{72}-{\mathcal
  H}(\bfm{p};\bfm{\mu},\bfm{\De},\bfm{M}).
\eea
Here we have defined
\beq
  \Ga_0=\left(\begin{array}{cc}
    \ga_0\otimes{\bf 1}_{9} & {\bf 0} \\
    {\bf 0} & {\bf 1}_{36}\\
  \end{array}
\right),\quad
\bar{\Ga}_0=\left(\begin{array}{cc}
    {\bf 1}_{36} & {\bf 0} \\
    {\bf 0} & \ga_0\otimes{\bf 1}_{9}\\
  \end{array}
\right).
\eeq
${\bf 1}_{72}$ denotes the $72$-dimensional unit matrix and 
${\mathcal H}$ is Hamiltonian density which is also $72\times72$ matrix
in the color, flavor and spinor space. 
In the same manner as shown in \cite{Ruster:2005jc}, we can lift the
spin degeneracy away from ${\mathcal H}$ as ${\mathcal H}={\mathcal
H}_{36}(p)\otimes P_+\oplus{\mathcal H}_{36}(-p)\otimes P_-$ with
$P_\pm\equiv\frac{1\pm\hat{p}\cdot\bfsm{\si}}{2}$ being the helicity
projectors.
${\mathcal H}_{36}(p)$ has a block-diagnalized form like
${\mathcal H}_{36}(p)={\mathcal H}_{12}^{(ur,dg,sb)}%
\oplus{\mathcal H}_{8}^{(bd,sg)}%
\oplus{\mathcal H}_{8}^{(sr,ub)}%
\oplus{\mathcal H}_{8}^{(ug,dr)}%
$.
Furthermore, the Nambu-Gor'kov degeneracy is manifest in the latter
three blocks; 
${\mathcal H}_{8}^{(\al,\be)}={\mathcal
H}_{4}^{(\al,\be)}\oplus\big[-{\mathcal
H}_{4}^{(\al,\be)}\big]$.
Thus we need to know only two matrix structures of ${\mathcal
H}_4^{(\al,\be)}\equiv{\mathcal H}_4^{\al\be}$ and 
${\mathcal H}_{12}^{(ur,dg,sb)}\equiv{\mathcal H}_{12}^{uds}$
for the evaluation of the effective potential.
We give the explicit form of these matrices below,
\beq
  {\mathcal H}^{\al\be}_{4}=\left( \begin{array}{cccc}
   M_\al-\mu_\al& p  & 0 & -i\De^{\al\be}\\
   p & -M_\al-\mu_\al  & -i\De^{\al\be} & 0\\
   0 & i\De^{\al\be}  & M_\be+\mu_\be &-p\\
   i\De^{\al\be}& 0 & -p& -M_\be+\mu_\be 
  \end{array}\right),
\eeq
where $\De^{\al\be}=\De_1, \De_2$ and $\De_3$ for
$(\al,\be)=(bd,sg), (sr,ub)$ and $(ug,dr)$, respectively, and
\begin{widetext}
\small
\beq
{\mathcal H}^{uds}_{12}=
\small
\left(%
\begin{array}{cccccccccccc}
   \!\!\!M_u\!-\!\mu_{ur} \!\!\!\!& p & 0 & 0 & 0 & 0 & 0 & -i\De_3 & 0 & 0 & 0 & -i \De_2
    \\ 
   p & \!\!\!\!-M_u\!-\!\mu_{ur}\!\!\!\!& 0 & 0 & 0 & 0 & -i\De_3 & 0 & 0 & 0 & -i\De_2 &
   0 \\ 
   0 & 0 & \!\!\!\!M_u\!+\!\mu_{ur}\!\!\!\! & -p & 0 & i\De_3 & 0 & 0 & 0 & i \De_2 & 0 & 0
   \\ 
   0 & 0 & -p & \!\!\!\!-M_u\!+\!\mu_{ur}\!\!\!\! & i\De_3 & 0 & 0 & 0 & i\De_2 & 0 & 0
   & 0 \\ 
   0 & 0 & 0 & -i \De_3 &\!\!\!\! M_d\!-\!\mu_{dg}\!\!\!\! & p & 0 & 0 & 0 & 0 & 0 & -i\De_1
   \\ 
   0 & 0 & -i \De_3 & 0 & p & \!\!\!\!-M_d\!-\!\mu_{dg}\!\!\!\! & 0 & 0 & 0 & 0 & -i\De_1
   & 0 \\ 
   0 & i \De_3 & 0 & 0 & 0 & 0 & \!\!\!\!M_d\!+\!\mu_{dg}\!\!\!\! & -p & 0 & i\De_1 & 0 &
   0 \\ 
   i\De_3 & 0 & 0 & 0 & 0 & 0 & -p & \!\!\!\!-M_d\!+\!\mu_{dg}\!\!\!\! & i\De_1 & 0 & 0
   & 0 \\ 0 & 0 & 0 & -i\De_2 & 0 & 0 & 0 & -i \De_1 & \!\!\!\!\Ms\!-\!\mu_{sb}\!\!\!\! & p
   & 0 & 0 \\ 
   0 & 0 & -i\De_2 & 0 & 0 & 0 & -i\De_1 & 0 & p & \!\!\!\!-\Ms\!-\!\mu_{sb}\!\!\!\! & 0 &
   0 \\ 
   0 & i\De_2 & 0 & 0 & 0 & i \De_1 & 0 & 0 & 0 & 0 & \!\!\!\!\Ms\!+\!\mu_{sb}\!\!\!\! &
   -p \\ 
   i\De_2 & 0 & 0 & 0 & i\De_1 & 0 & 0 & 0 & 0 & 0 & -p & \!\!\!\!-{\Ms}\!+\!\mu_{sb}\!\!\! \\
\end{array}\right).\label{eq:12matrix}
\eeq
\end{widetext}
Because the Nambu-Gor'kov degeneracy is not removed, the eigenvalues 
of this matrix have six sets of the doublet $(\eps,-\eps)$
corresponding to the energies for a quasi-quark and its Nambu-Gor'kov
partner (anti-quasi-quark).

Let us write $18(=72\div2\div2)$ independent eigenvalues of ${\mathcal
H}$ as $\{\eps_\al\}$ with the index $\al$ redefined to the label for
the quasi-quarks ($\alpha=1,2,\cdots,18$) from that for the
direct products of color and flavor ($\alpha=ur,dg,sb,db,sg,sr,ub,ug,dr$).
Then the thermodynamic potential can be simplified to
\bea
  \Om_q&=&\frac{4}{G_d}\sum\De_\eta^2+\frac{N_c}{G_s}\sum(M_i-m_i)^2\nn
        & &-2\sum_{\al=1}^{18}\int\frac{d\bfm{p}}{(2\pi)^3}%
	   \left[\frac{|\eps_\al|}{2}%
           +T\log\left(1+e^{-|\eps_\al|/T}\right)\right]\nn
	& &+2\sum_{\al=1}^{18}%
	   \int\frac{d\bfm{p}}{(2\pi)^3}\frac{p}{2}.\nonumber
\eea
We have subtracted the vacuum contribution in the system with the nine
massless quarks.
It is difficult to numerically search the minima with respecting 
the neutrality constraints from this effective potential alone.
Therefore, we search them with the aid of the gap equation which is the
$\De$-derivative of the effective potential.
If possible, numerical derivatives should be avoided because
the numerical errors associated with them are not well controllable.
To avoid them, we express the gap equations in terms of the 
eigenvectors (eigen-spinors) of ${\mathcal H}$. First, we simply
differentiate \Eqn{det} with respect to $\De_\eta$ and equating the
result to zero to obtain,
\beq
  \frac{8}{G_d}\De_\eta%
  =-T\sum_n\int\frac{d\bfm{p}}{(2\pi)^3}\tr\left[%
  \frac{1}{i\om_n{\bf 1}_{36}-{\mathcal H}_{36}}%
  \frac{\p{\mathcal H}_{36}}{\p\De_\eta}\right].\label{der1}
\eeq
Here ${\mathcal H}_{36}$ is the reduced Nambu-Gor'kov Hamiltonian
density defined by removing the spin degeneracy from ${\mathcal H}$
as was introduced above.
Also we note that the reduced Hamiltonian density takes the form
\beq
\ba{rcl}
 {\mathcal H}_{36}&=&{\mathcal H}_0 + M_u\bfm{\phi}_{M_u} +
   M_d\bfm{\phi}_{M_d} + \Ms\bfm{\phi}_{\Ms}\\[1ex]
 &&+ \De_1\bfm{\phi}_1 + \De_2\bfm{\phi}_2+\De_3\bfm{\phi}_3\\[1ex]
 &&-3\mu\bfm{B}-\mue\bfm{Q}_e-\mu_3\bfm{T}_3-\mu_8\bfm{T}_8,
\ea
\eeq
with ${\mathcal H}_0={\mathcal
H}_{36}|_{\bfsm{\mu}=\bfsm{0},\bfsm{\De}=\bfsm{0},\bfsm{M}=\bfsm{0}}$
being the Hamiltonian density for the system with nine free massless
quarks.
$\bfm{\phi}_{M_{i}}, \bfm{\phi}_{\De_\eta}$ and $\bfm{B}, \bfm{Q}_e,
\bfm{T}_{3,8}$ are the $36\times 36$ matrices with constant matrix
elements.
These matrices can be obtained by differentiating the reduced
Hamiltonian matrix with respect to
$M_{u,d},\,\Ms,\,\De_{1,2,3},\,\mu,\,\mu_{e,3,8}$ as follows.
\beq
\ba{l}
 \bfm{\phi}_{M_{u,d}}=\frac{\p {\mathcal H}_{36}}{\p M_{u,d}},\quad%
 \bfm{\phi}_{\Ms}=\frac{\p {\mathcal H}_{36}}{\p \Ms},\\[1ex]
 \bfm{\phi}_{1}=\frac{\p {\mathcal H}_{36}}{\p\De_1},\quad%
 \bfm{\phi}_{2}=\frac{\p {\mathcal H}_{36}}{\p\De_2},\quad%
 \bfm{\phi}_{3}=\frac{\p {\mathcal H}_{36}}{\p\De_3},\\[1ex]
 \bfm{B}=-\frac{1}{3}\frac{\p {\mathcal H}_{36}}{\p\mu},\quad%
 \bfm{Q}_e=-\frac{\p {\mathcal H}_{36}}{\p\mue},\\[1ex]%
 \bfm{T}_3=-\frac{\p {\mathcal H}_{36}}{\p\mu_3},\quad
 \bfm{T}_8=-\frac{\p {\mathcal H}_{36}}{\p\mu_8}.\\[1ex]
\ea
\eeq

Because ${\mathcal H}_{36}$ is an Hermitian matrix for any momentum,
we can define the complete set of spinors $\{|p,\al,\pm\gt\}$ by
the eigenvalue equation
\beq
  {\mathcal H}_{36}|p,\al,\pm\gt%
  =\pm\eps_{\al}(p)|p,\al,\pm\gt,\label{eigenvectors}
\eeq
$\al=1,\cdots,18$ and $\pm$ denote the Nambu-Gor'kov charges.
Then we have
\beq
  \lt p,\al,\pm|\bfm{\phi}_\eta|p,\al,\pm\gt=%
  \pm\frac{\p\eps_\al}{\p\De_\eta}, \quad (\eta =1,2,3).
\eeq
Thus we find that \Eqn{der1} is reduced to 
\beq
  \frac{8}{G_d}\De_\eta
  =\sum_{\al=1}^{18}\int\!\!\!\frac{d\bfm{p}}{(2\pi)^3}%
  \tanh\left(\frac{\eps_\al}{2T}\right)%
  \lt p,\al,+|\bfm{\phi}_\eta|p,\al,+\gt,\nonumber
\eeq
where the Matsubara summation has been performed.
In much the same way, we obtain
\beq
\ba{l}
  \dsp\frac{2N_c}{G_s}(M_i-m_i)\\[1ex]
  \dsp\qquad\quad=\sum_{\al=1}^{18}\int\!\!\!\frac{d\bfm{p}}{(2\pi)^3}%
  \tanh\left(\frac{\eps_\al}{2T}\right)%
  \lt p,\al,+|\bfm{\phi}_{M_i}|p,\al,+\gt.\\[1ex]
\nonumber
\ea
\eeq
These formulae enable us to evaluate the gap
equation solely by computing the $18$-dimensional eigen-spinors
defined by \Eqn{eigenvectors} and some matrix elements in these 
bases \cite{Fredrik}
without recourse to numerical derivatives done in
\cite{Alford:2003fq,Fukushima:2004zq,Ruster:2005jc}.

The charge neutrality constraints can  be also expressed 
in terms of the matrix elements in the basis composed of these eigen-spinors. 
A straightforward application of the above method to
$\rho_{3,8}=-\p\Om_{q}/\p\mu_{3,8}=0$,
$\rho_e^q=-\p\Om_{q}/\p\mue=-n_e$ and 
$\rho_B=-\frac{1}{3}\p\Om_{q}/\p\mu$ leads
\bea
 0&=&\sum_{\al=1}^{18}%
             \int\!\!\!\frac{d\bfm{p}}{(2\pi)^3}%
	     \left(2f_F(\eps_\al)-1\right)%
	     \lt p,\al,+|\bfm{Q}_{3,8}%
	     |p,\al,+\gt,\nn
 -n_e&=&\sum_{\al=1}^{18}%
             \int\!\!\!\frac{d\bfm{p}}{(2\pi)^3}%
	     \left(2f_F(\eps_\al)-1\right)%
	     \lt p,\al,+|\bfm{Q}_{e}%
	     |p,\al,+\gt,\nn
 \rho_B&=&\sum_{\al=1}^{18}%
             \int\!\!\!\frac{d\bfm{p}}{(2\pi)^3}%
	     \left(2f_F(\eps_\al)-1\right)%
	     \lt p,\al,+|\bfm{B}%
	     |p,\al,+\gt.\nonumber
\eea
Here, the Fermi-Dirac distribution function
$f_F(\eps)=1/(e^{\eps/T}+1)$ is introduced.

At zero temperature, $f_F(\eps)=\th(-\eps)$ so that the net charge
will be accumulated in the {\em blocking region} where $\eps<0$.
For the neutrality constraints to be satisfied, 
there must be the opposite charge
density coming from a non-quark sector or from the quark sector with
the finite background charge density $\int\frac{d\bfsm{p}}{(2\pi)^3}%
\sum_\al\lt p,\al,+|\bfm{Q}|p,\al,+\gt\ne 0$,
which are supplied by tuning the charge chemical potentials
$\mue,\,\mu_3$ and $\mu_8$.

For a later convenience, we introduce here a charge generator
\beq
 \Qtilde=-Q-T_3-\frac{1}{2}T_8,
\eeq
the operation of which keeps the CFL state invariant (neutral)
 and thus represents an unbroken $U(1)$ symmetry in the CFL phase
 \cite{reviews}.
If we choose the orthogonal basis of the broken charges $(X,Y)$ in addition
to unbroken $\Qtilde$ as \cite{Alford:2002kj},
\beq
\ba{rcl}
 X&=&Q+T_3-4T_8,\\[3ex]
 Y&=&Q-T_3,\\[1ex]
\ea
\nonumber
\eeq
the chemical potentials for these charges become
\beq
\ba{rcl}
  \mu_{\Qtilde}&=&-\frac{4}{9}\bigl(\mue
  +\mu_3 + \half\mu_8\bigr),\\[3ex]
  \mu_X&=&\frac{1}{18}(-\mu_Q+\mu_3-4\mu_8),\\[3ex]
  \mu_Y&=&\frac{1}{2}(-\mu_Q-\mu_3).\\[1ex]
\ea
\nonumber
\eeq
We find that the matrix representation of $\Qtilde$ operator 
in the $32$-dimensional color-flavor mixed Nambu-Gor'kov base 
is given by 
\beq
\bfm{\Qtilde}=-\frac{\p{\mathcal H}_{36}}%
 {\p\mu_{\Qtilde}}=-\bfm{Q}_e-\bfm{T}_3-\frac{1}{2}\bfm{T}_8.
\eeq
Because this commutes with the Nambu-Gor'kov Hamiltonian density 
${\mathcal H}_{36}$ as it should be, the quasi-particles (eigen-spinors)
have definite $\Qtilde$-charges which can be shown to be of integers
$(+1,0,-1)$ \cite{reviews}.
It is also to be noted here, that thermodynamic potential in the quark
sector $\Om_{q}$ does not depend on $\mu_{\Qtilde}$ in the
fully gapped CFL phase so that it is a $\Qtilde$-insulator
\cite{Alford:2002kj}.
Because of this, the value of $\mu_{\Qtilde}$ in the CFL phase ($T=0$)
cannot be determined by the $\Qtilde$ neutrality condition in the quark
sector; it should be determined completely by the vanishing-point of
very gentle slope of potential curvature coming from the electron
sector ($\mue^3/3\pi^2=0$).

\vspace*{7ex}
\section{Numerical Results and Discussions}\label{Numericalresults}
\begin{table*}[tp]
 \begin{tabular}{|r||ccccc|c|cccc|}
  \hline
  \multicolumn{1}{|c||}{\bf }& \multicolumn{5}{c|}{\bf\sc Gap and
   mass} & {\bf\sc Conditions} & 
  \multicolumn{4}{c|}{\bf\sc Gapless quark 
   and $\mathstrut\Qtilde$ charge} \\
  \multicolumn{1}{|c||}{\bf\sc  }& 
  \multicolumn{5}{c|}{\bf\sc parameters }  
  & {\bf\sc  for}
  &~~($ur$-$dg$-$sb$)~~&~~($db$-$sg$)~~&~~($ub$-$sr$)~~&~~($ug$-$dr$)~~\\
  \multicolumn{1}{|c||}{\bf\em Phase}& $\De_1(ds)$ & $\De_2(us)$ &
  $\De_3(ud)$ & $M$ & $\Ms$ & {\bf\sc  chemical potentials}
  &~~$0$,~~$0$,~~$0$~~&~~$0$,~~$0$~~&~~$+1$,$-1$~~&~~$+1$,$-1$~~\\
  \hline
  CFL (9)$\frac{\mathstrut}{\mathstrut}$ 
  & $\De_1$ & $\De_2$ & $\De_3$ &
  & $\Ms$ &[$\mue=0$] &\multicolumn{4}{c|}{all quark modes are fully
  gapped}\\ \hline
  gCFL${}_8$ (8)$\frac{\mathstrut}{\mathstrut}$ 
  & $\De_1$ & $\De_2$ & $\De_3$
  & & $\Ms$ & $\de\mu_{dbsg}%
      +\frac{\Ms^2}{4\mu}\agt\De_{1}$ & & $db$ &   & \\ \hline
  gCFL (7)$\frac{\mathstrut}{\mathstrut}$ 
  & $\De_1$ & $\De_2$ & $\De_3$
  & & $\Ms$ & $\de\mu_{dbsg(ubsr)}%
      +\frac{\Ms^2}{4\mu}\agt\De_{1(2)}$ & & $db$ & $ub$  & \\ \hline
  uSC (6)$\frac{\mathstrut}{\mathstrut}$ 
  & & $\De_2$ & $\De_3$ & & $\Ms$
  & [$\mue=0$] & $dg$-$sb$ (1) &($db$,~$sg$) & & \\ \hline
  guSC (5)$\frac{\mathstrut}{\mathstrut}$ 
  & & $\De_2$ & $\De_3$ & &
  $\Ms$ & $\de\mu_{ubsr}+\frac{\Ms^2}{4\mu}\agt\De_2$  &
  $dg$-$sb$ (1) &  ($db$,~$sg$) & $ub$ & \\ \hline\hline
  2SC (4)$\frac{\mathstrut}{\mathstrut}$ 
  & & & $\De_3$ & & $\Ms$ &
  [$\mu_3=0] $ & $sb$ & ($db$,~$sg$) & ($ub$,~$sr$) & \\ \hline
  g2SC (2)$\frac{\mathstrut}{\mathstrut}$ 
  & & & $\De_3$ & & $\Ms$ &
  [$\mu_3=0$], $\de\mu_{dgur}=\de\mu_{drug}>\De_3$ &
  $dg$,~$sb$ & ($db$,~$sg$) & ($ub$,~$sr$) & $dr$\\ \hline
  dSC (6)$\frac{\mathstrut}{\mathstrut}$ 
  & $\De_1$ & & $\De_3$ & & $\Ms$
  & & $ur$-$sb$ (1) & & ($ub$,~$sr$) & \\ \hline
  gdSC (5)$\frac{\mathstrut}{\mathstrut}$ 
  & $\De_1$ & & $\De_3$ & &
  $\Ms$ & $\de\mu_{dbsg}+\frac{\Ms^2}{4\mu}\agt\De_1$ &
  $ur$-$sb$ (1)& $db$ & ($ub$,~$sr$) & \\ \hline\hline
  2SCus (4)$\frac{\mathstrut}{\mathstrut}$ 
  & & $\De_2$ & & & $\Ms$ & &
  $dg$  & ($db$,~$sg$)&  & ($ub$,~$sr$) \\ \hline\hline
  UQM (0)$\frac{\mathstrut}{\mathstrut}$ 
  & & & & & $\Ms$ &
  [$\mu_3=\mu_8=0]$ &\multicolumn{4}{c|}{all quarks are ungapped.} \\
  \hline
  $\chi$SB (0)$\frac{\mathstrut}{\mathstrut}$ 
  & & & &$M$ &$\Ms$ &
  [$\mu_3=\mu_8=0]$ & \multicolumn{4}{c|}{all quarks are massive.} \\ 
  \hline
 \end{tabular}
 \caption[]{%
{The nonzero gap parameters, some conditions between gaps and chemical
 potentials, and the gapless quarks in each phase.
 The figure in the parenthesis in the first column represents the number
 of gapped quasi-quark mode. 
 We have defined the ($ai$-$bj$) relative chemical potential by
 $\mu_{aibj}\equiv(\mu_{ai}-\mu_{bj})/2$.
 ``$dg$-$sb$ (1)'' means that one of the linear combinations, the $dg$
 quark and the $sb$ hole remains gapless. 
 The equation for chemical potentials in a bracket must necessarily
 hold for some symmetry or kinetic reason.}
 }
 \label{gaps}
\end{table*}

In this section, we present our numerical results 
of the solution of the gap equations, and provide the phase diagrams
in the $(\mu,T)$-plane for several values of diquark coupling $G_d$.

Before that, we fix our model parameters.
We take the chiral SU$(2)$ limit for the $u$, $d$ current quark masses
($m_u=m_d=0$) and $\ms=80\MeV$ \cite{Abuki:2004zk}.
These values might slightly underestimate the effect of the current
masses because $m_{u,d}(2\GeV)=3$-$4\MeV$ and $\ms(2\GeV)=80$-$100\MeV$
according to the full lattice QCD simulation \cite{AliKhan:2001tx}.
Also we restrict the variational space by putting $M=M_u=M_d$ for
simplicity.
This simplification does not matter in the chiral symmetry restored
phase \cite{Ruster:2005jc,Blaschke:2005uj}.

For comparison with the previous work \cite{Ruster:2005jc}, we write
down the dimensionless parameters adopted in this study; 
\beq
\ba{c}
  \dsp m_{u,d}/\Lam=0,\quad m_{s}/\Lam=0.1,\\[2ex]
  \dsp G_S\Lam^2\equiv \frac{G_s \Lam^2}{8N_c}=2.17.\\[1ex]
\ea
\nonumber
\eeq
The value of $G_s$ is chosen so that the dynamical quark mass at $\mu=0$
is $400\MeV$ for the cutoff $\Lam=800\MeV$ just for comparison with our
previous work \cite{Abuki:2004zk}.
Note, however, that this coupling is a little larger than the value
extracted in the NJL model analysis of the meson spectroscopy with
instanton induced six-quark coupling \cite{Hatsuda:1994pi}, \ie,
$G_S\Lam^2=1.835$ which is adopted in \cite{Ruster:2005jc}.
We shall perform the calculation with following five different
values of the diquark coupling ($G_D\equiv G_d/16$): 
\bi
\item[1.]{\bf\em extremely weak coupling:}\\
	 $\De_0=25\MeV\leftrightarrow G_D\Lam^2=0.91\,\,(G_D/G_S=0.42)$
\item[2.]{\bf\em weak coupling:}\\
	 $\De_0=80\MeV\leftrightarrow G_D\Lam^2=1.37\,\,(G_D/G_S=0.63)$
\item[3.]{\bf\em intermediate coupling:}\\
	 $\De_0=125\MeV\leftrightarrow G_D\Lam^2=1.69\,\,(G_D/G_S=0.78)$
\item[4.]{\bf\em strong coupling:}\\
	 $\De_0=160\MeV\leftrightarrow G_D\Lam^2=1.92\,\,(G_D/G_S=0.88)$
\item[5.]{\bf\em extremely strong coupling:}\\
	 $\De_0=200\MeV\leftrightarrow G_D\Lam^2=2.17\,\,(G_D/G_S=1.00)$
\ei
The values of $G_D/G_S$ for the intermediate and extremely strong
coupling cases are similar to those employed in \cite{Ruster:2005jc},
\ie, $G_D/G_S=3/4$ and $G_D/G_S=1$.
However, the following notice is in order here.
(i)
Our strange quark mass
$\ms/\Lam=0.1$ is about one half of the value $\ms/\Lam=0.23$
adopted in \cite{Ruster:2005jc}.
(ii)
We did not included the six-quark interaction which effectively
increases the scalar coupling $G_S$.
These two differences make our case favorable to the pairing
phases rather than the {\em unpaired quark matter} (UQM) phase or the
{\em chiral-symmetry broken} ($\chi$SB) phase.
In fact, we will see that the phase diagrams for our weak coupling
and intermediate coupling cases seem more or less to correspond to
those for $G_D/G_S=3/4$ and $G_D/G_S=1$ in \cite{Ruster:2005jc},
respectively.

We consider the candidates of phase listed in TABLE~\ref{gaps} as in
\cite{Abuki:2004zk} in the numerical analyses below.

\begin{figure}[htp]
  \includegraphics[width=0.42\textwidth,clip]{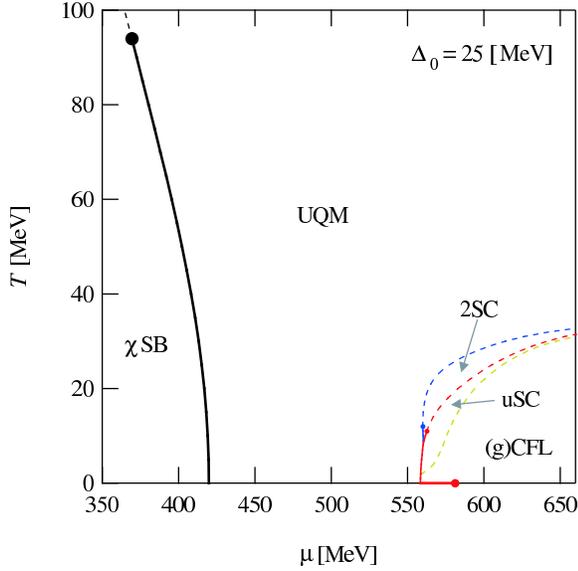}
  \caption[]{The phase diagram for the extremely weak coupling
 ($\De_0=25\MeV$).
  Sold (dashed) lines mean first (second) order transitions.
  Large dot placed on the $\mu$-axis ($\mu=581.1\MeV$) represents the
  transition point at which the CFL phase on the high density side
  continuously turns into the gCFL phase on the low density side.}
  \label{phase_weak}
\end{figure}
\subsection{Phases for extremely weak coupling}\label{extweak}
Let us first discuss the extremely weak coupling case ($\De_0=25\MeV$).
The phase structure in this case is displayed in Fig.~\ref{phase_weak}.
At a first glance, we notice that the UQM phase without any symmetry
breaking dominates the phase diagram pushing the superconducting phases
to the high density regime. 
This is because the energy gain due to the $\lt \bar{s}s\gt$
condensation in the UQM phase is larger than the paring energy 
under the stress as is clarified in \cite{Abuki:2004zk}.
There are also several thermal phase transitions.
The thermally-robustest pairing phase is the 2SC and the second phase
in this case is the uSC as is found in
\cite{Ruster:2005jc,Blaschke:2005uj}.
In the following, we shall discuss the features of
these phase transitions in detail.
We first make a close examination on the zero temperature case, 
and then investigate the finite temperature case.

\begin{figure}[htp]
  \includegraphics[width=0.42\textwidth,clip]{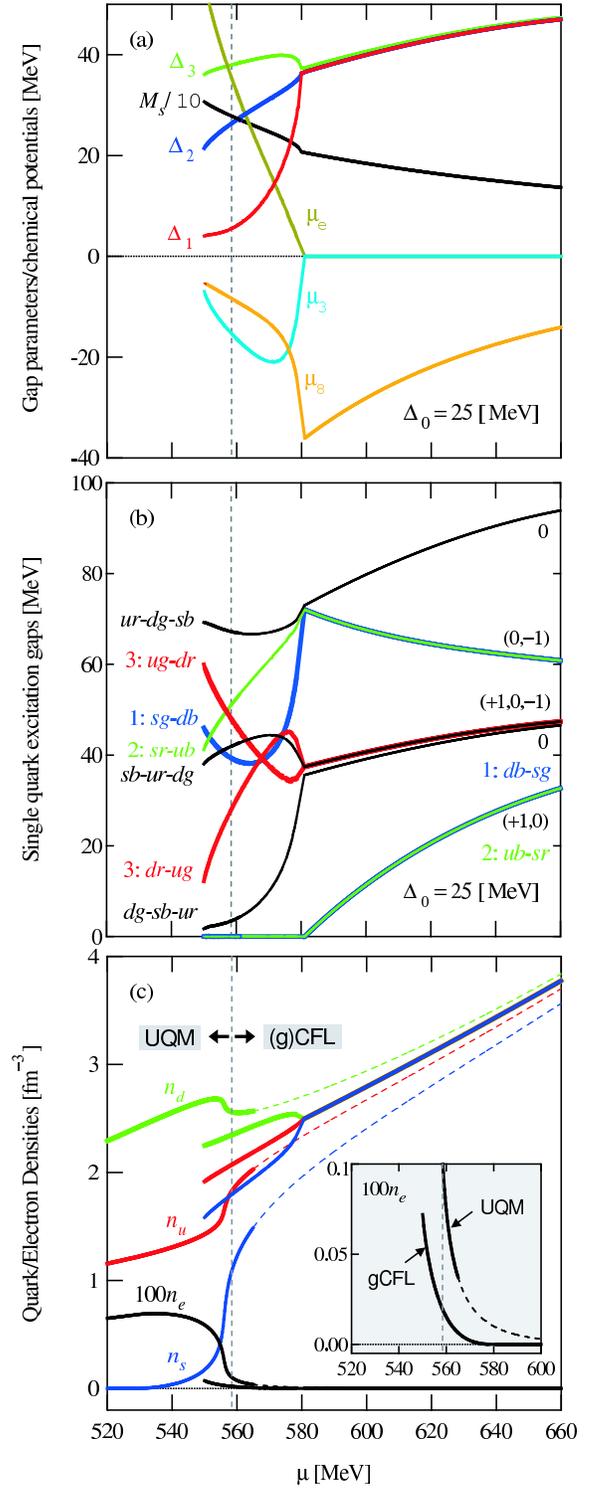}
  \caption[]{%
  {\bf(a)}~~The (g)CFL solution of the gap equation 
   under the neutrality constraints. 
   The (g)CFL is energetically taken over by 
   the UQM phase on the left side of $\mu=558.4\MeV$
   denoted by the vertical dashed line.
  {\bf (b)}~~The excitation gaps in the nine quasi-quark spectra 
   as a function of $\mu$. 
   The $\Qtilde$-charge associated with each mode is also indicated.
  {\bf (c)}~~The quark and electron densities are displayed. 
   The lines on the left hand side are densities in the UQM phase,
   while those on the right hand side are densities along the (g)CFL
   solution. 
  }
  \label{solution_weak}
\end{figure}
\setcounter{paragraph}{0}
\vspace*{0.3cm}
\paragraph{CFL/gCFL and gCFL/UQM transitions at $T=0$:}
The phases realized at $T=0$ are the CFL, gCFL and UQM states.
In Fig.~\ref{solution_weak}(a), we show the gaps, masses, and
chemical potentials along the (g)CFL solution of the gap equation.
For $\mu>558.4\MeV$
we confirm that the condensation energy in the (g)CFL solution 
is largest among those for all the candidates.
The excitation gaps of the nine quasi-quarks in the CFL phase are shown
 in FIG.~\ref{solution_weak}(b).
We can see from the figure that one quasi-quark has the largest
excitation gap, and other quarks have relatively small gaps.
The latter eight modes are the remnants of the color-flavor
octet modes in the pure CFL phase at $\Ms=0$.
In the CFL phase, the non-vanishing $\mu_8$ and $\Ms$ are {\em
dynamically} realized so that the original symmetry of color-flavor 
diagonal SU$(3)_{C+V}$ is {\em explicitly} broken down to color 
SU$(2)_{C+V}$ (color-flavor isospin) in the {\em Lagrangian level};
because of this, the octet modes split into the isospin-singlet 
mode (like eta) and two set of doublet (kaonic) modes and the 
triplet (pionic) modes as
$8\to3\,(+1,0,-1)\,+\,2\,(+1,0)\,+\,2\,(0,-1)\,+\,1\,(0)$
where the associated $\Qtilde$-charge is indicated in the parenthesis. 
One of the most striking features of the CFL phase is the absence
of electrons; the electric neutrality is realized solely by the quark
sector $n_u=n_d=n_s$ as can be seen in Fig.~\ref{solution_weak}(c).
This CFL phase behaves as $\Qtilde$-insulator because of the 
absence of gapless $\Qtilde$-carriers \cite{Alford:2003fq}.

As the density is decreased, the stress energy
$-\mu_8(\mu)/2+\Ms^2(\mu)/4\mu$ \cite{Alford:2003fq} becomes
large in the CFL state. 
When it reaches $\De_1=\De_2$, the first qualitative change takes place;
this happens when the chemical potential is decreased 
down to $\mu=581.1\MeV$.
At this point, the CFL state continuously turns into 
the gCFL phase.
Just at the gCFL onset, the excitation gaps for the one 
doublet modes reaches zero as can be seen 
in Fig.~\ref{solution_weak}(b).
The situation is similar to the $K^{0}$ condensation (with a small
fraction of $K^+$ condensation) \cite{Kaplan:1986yq,Bedaque:2001je}
because the $ub$-$sr$ and $db$-$sg$ modes belong to the 
color-isospin SU$(2)_{C+V}$ doublet, and have $\Qtilde=+1$ and $0$,
respectively.

When the chemical potential is decreased further,
the gap parameters split into three different values
although they are all still finite. 
Accordingly, the isospin SU$(2)_{C+V}$ symmetry gets broken by $\mue$
and $\mu_3$ so that the gaps in quasi-quark
dispersions all take different values (see FIG.~\ref{solution_weak}(b)).
We remark that the electron chemical potential or its density actually
serves as an order parameter of the CFL/gCFL (insulator/metal)
transition as claimed in \cite{Alford:2003fq}.

The gCFL phase continues to be the ground state down to 
$\mu=558.4\MeV$ below which the UQM phase is more favorable
in terms of the thermodynamic potential.
When the transition gCFL $\to$ UQM takes place, there should be
the large re-configuration of the flavor contents as can be
seen in FIG.~\ref{solution_weak}(c).
Thus, this transition requires a lot of electro-weak processes
which include the $d$ production like $ug(ur)\to dg(dr)+e^++\nu_e$
in addition to the decay to the gapless modes accompanied
by  the electron
production $db\to ub+e^-+\bar{\nu}_e$ and the $s$ quark 
decay $sb+u\to db+u$.
We should note that there still remains an open interesting 
question how the UQM droplets are dynamically formed in the gCFL phase
and grow against the surface tension.

We have seen that, as the density is decreased, the CFL phase turns into
the gCFL phase, and then the gCFL phase gets taken over by the UQM phase.
Accordingly, the number of gapped quasi-quark modes decrease as 9 (CFL)
$\to$ 7 (gCFL) $\to$ 0 (UQM) at $T=0$ as shown in the previous work
\cite{Abuki:2004zk}.
Next, we will discuss how the situation is changed in the $T\ne0$ case.

\begin{figure}[thp]
  \includegraphics[width=0.42\textwidth,clip]{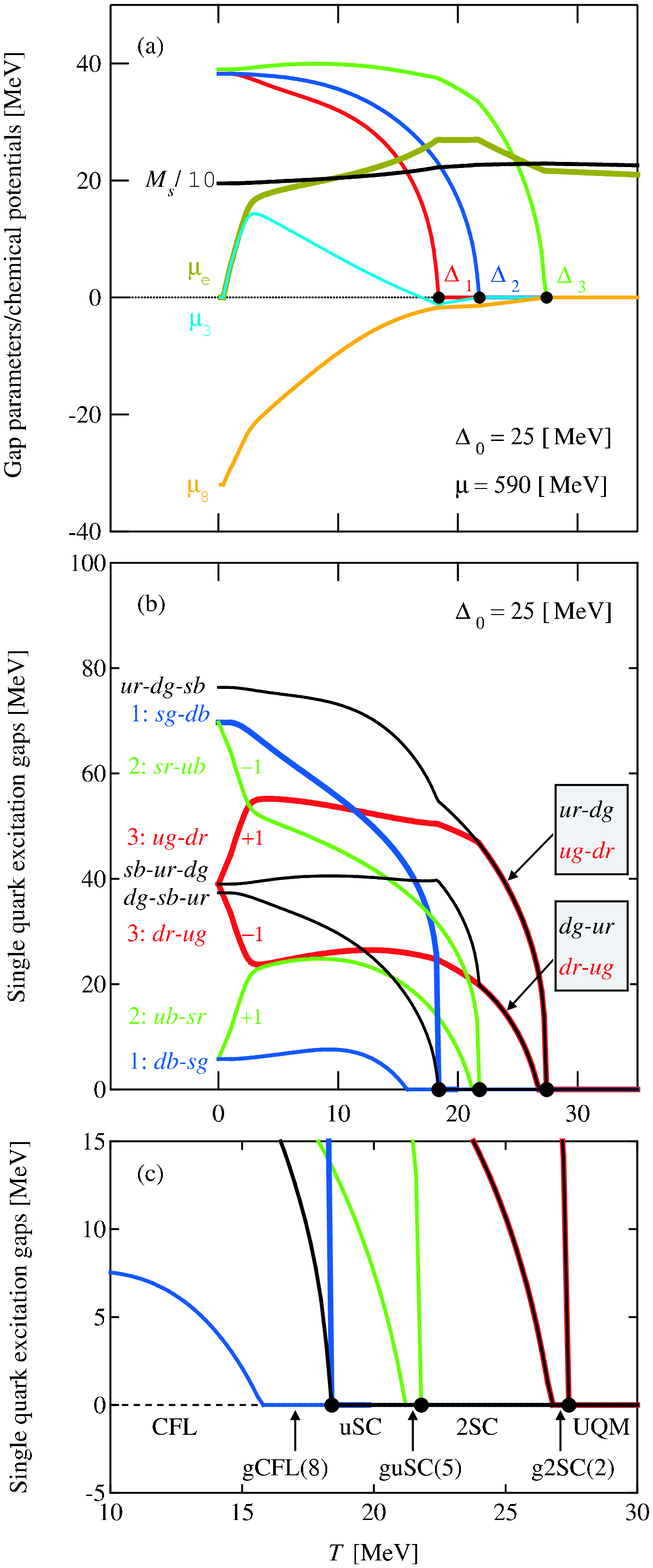}
  \caption[]{{\bf (a)}~~The gap and mass parameters, and the chemical
  potentials as a function of $T$ at $\mu=590\MeV$ so that the system at
  $T=0$ stays in the CFL phase (see FIG.~\ref{solution_weak}).
  {\bf (b)}~~The nine quasi-quark gaps versus $T$. 
  Nonzero $\Qtilde$-charge is also indicated by $\pm1$.
  {\bf (c)}~~Just an enlargement of the figure (b).
  }
  \label{gapT_weak}
\end{figure}

\begin{figure*}[thp]
 \begin{minipage}{0.42\textwidth}
  \includegraphics[width=0.9\textwidth,clip]{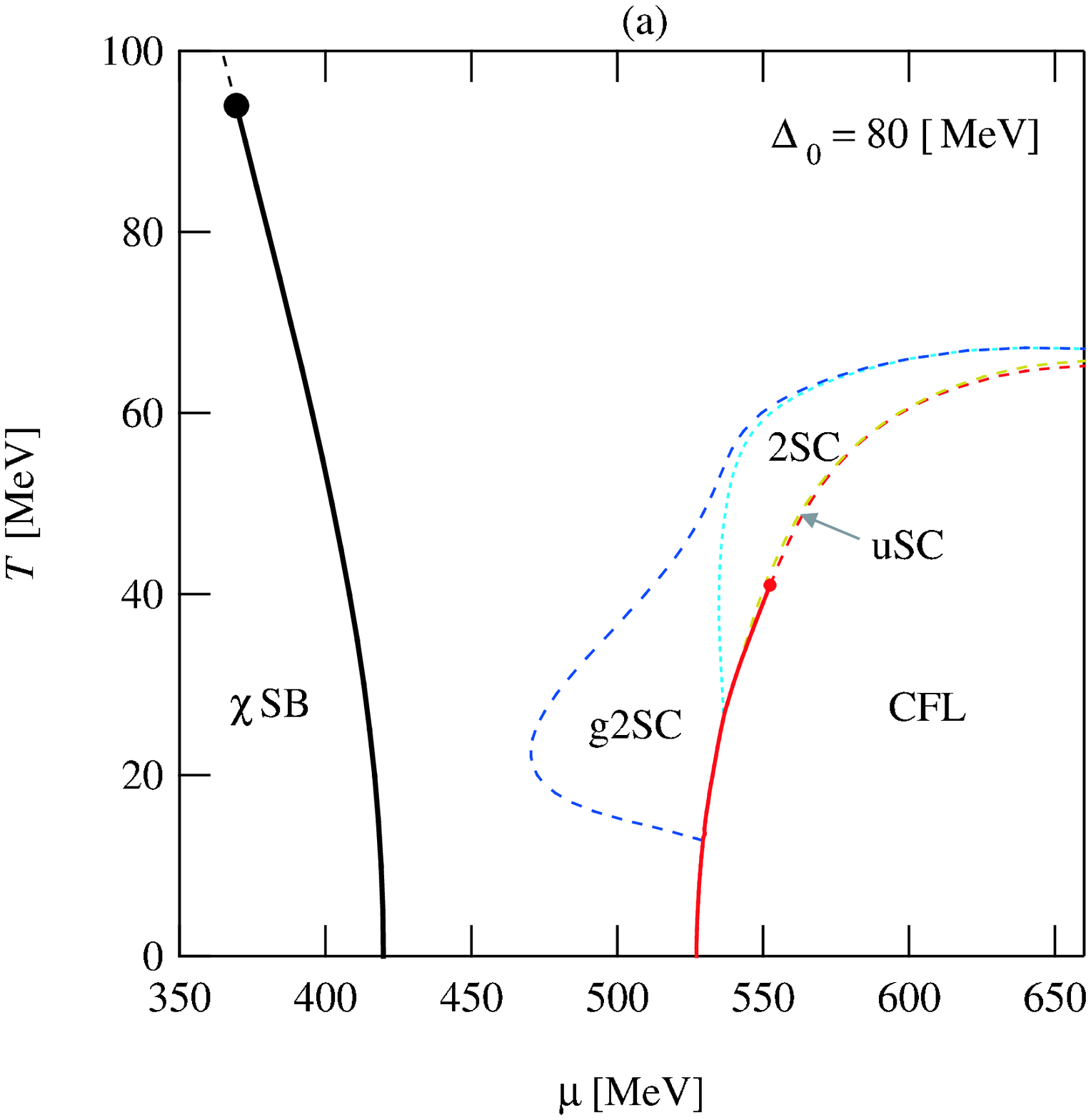}
 \end{minipage}%
 \hfill%
 \begin{minipage}{0.42\textwidth}
  \includegraphics[width=0.9\textwidth,clip]{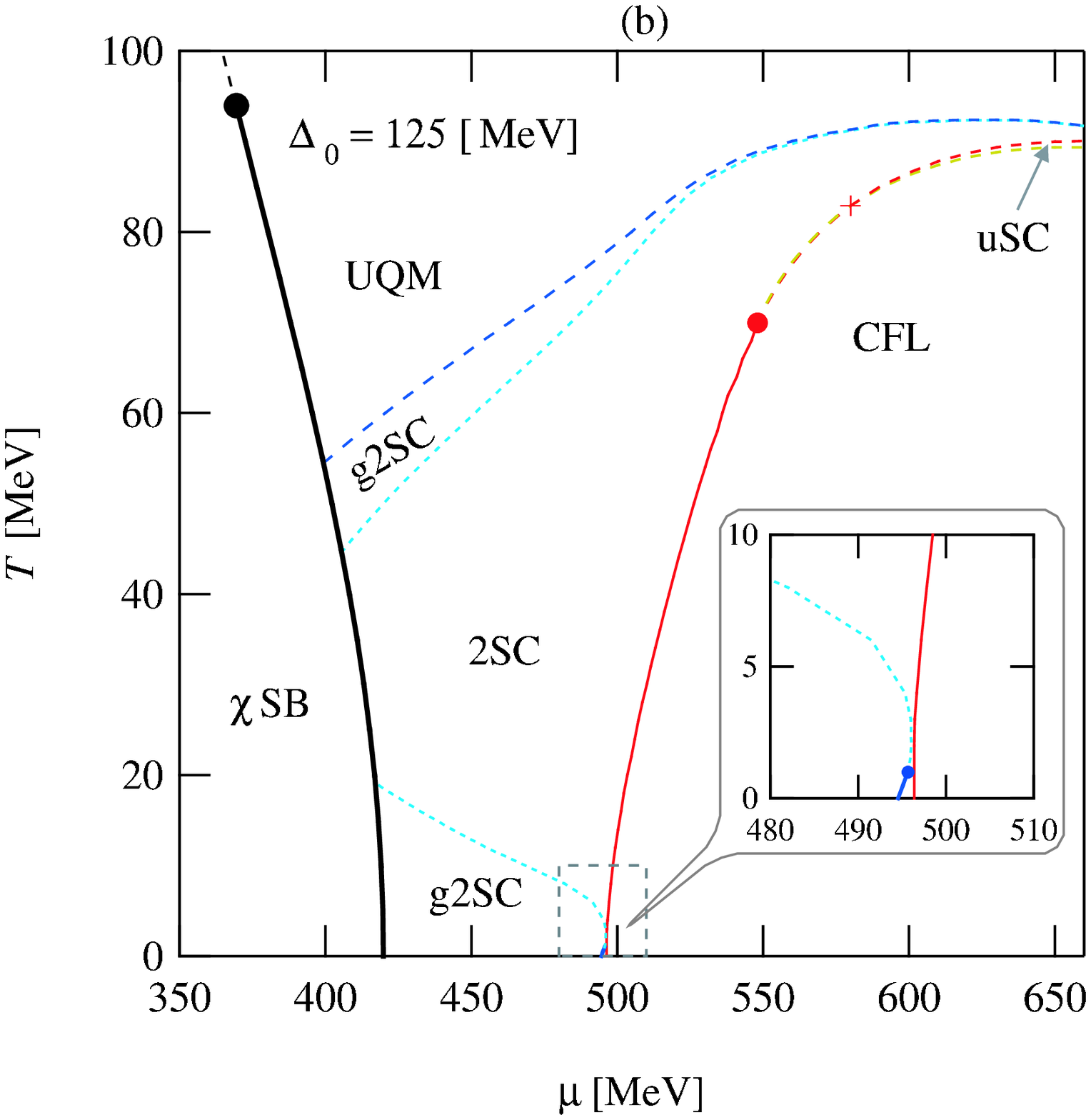}
 \end{minipage}
 \caption[]{The phase diagram for the weak coupling $(\De_0=80\MeV)$
 {\bf (a)}, and for the intermediate coupling $(\De_0=125\MeV)$ {\bf
 (b)}.}
 \label{phase_int}
\end{figure*}

\vspace*{0.3cm}
\paragraph{Phase transitions and crossovers for finite temperature:}
We show here that the quark matter undergoes a sequence of the 
transitions,
CFL(9) $\to$ gCFL${}_8(8)$ $\to$ uSC(6) $\to$ guSC(5) $\to$ 2SC(4) $\to$
g2SC(2) $\to$ UQM(0) as $T$ becomes large;
the number of the gapped modes in each phase is indicated with
a parenthesis.

In FIG.~\ref{gapT_weak}(a), we show the $T$-dependence of the gaps,
masses, and chemical potentials. 
The figure shows that the $\De_1$ first melts and then $\De_2$ disappears 
as $T$ is increased.
The thermally-robustest pairing is the 2SC, while second one is the uSC,
which is in agreement with the result in \cite{Ruster:2005jc}.
This conclusion will be found to be also consistent with the
Ginzburg-Landau analysis given in Sec.~\ref{GLtheory}, where we extend
the previous study \cite{Iida:2003cc} and see that the large value of
the strange quark mass actually disfavors the dSC phase.

In order to study the thermal transition in more detail, we have 
also plotted the $T$-dependence of the gaps in the nine quasi-quark
spectra in FIG.~\ref{gapT_weak}(b). 
FIG.~\ref{gapT_weak}(c) is just an enlargement of (b). 
The three large points on the horizontal line 
indicate $T=T_{c\eta}$ $(\eta=1,2,3)$ at which
$\De_\eta$ vanishes (the same as the large points in
FIG.~\ref{gapT_weak}(a)).
When the temperature is increased from $T=0$, the first qualitative
change occurs at $T\sim15.8\MeV$ where one of the $db$-$sg$
dispersion becomes gapless and the gCFL${}_8$ phase (gCFL in
\cite{Ruster:2005ib}) sets in.
Unlike the gCFL phase at $T=0$, the $ub$-$sr$ dispersion does not
become gapless at this point.
From the viewpoint of the insulator-to-metal transition, however, the
sharp phase transition is smoothen not only due to the absence of the
gapless $ub$-$sr$ mode with $\Qtilde=1$, but also to the
thermally-excited on-shell quasi-quarks at finite temperature
as the latter ingredient is already noticed in \cite{Fukushima:2004zq}.
In fact, from the FIG.~\ref{gapT_weak}(b), we can see that the 
$ub$-$sr$ ($\Qtilde=+1$) mode is lighter than the $db$-$sg$
($\Qtilde=-1$) mode for $T\alt5\MeV$ and accordingly there is a little
excess of the quasi-quarks with $\Qtilde=+1$ in the system.
In order to achieve the neutrality, there must be the equal amount of
electrons so that $\mue$ takes positive finite value as long as $T$ is
finite.
As a result, the CFL-gCFL${}_8$ (insulator-metal) transition becomes
a smooth crossover.

When $T$ is increased further beyond the gCFL${}_8$ onset, $\De_1$
disappears at $T\sim18.4\MeV$.
This is a second order phase transition of gCFL${}_8$ $\to$ uSC.
As a consequence, two quasi-quark modes become gapless as is seen in
FIG.~\ref{gapT_weak}(b);
one is the $sg$-$db$ mode \ie, the partner of the gapless $bd$-$sg$
mode, while the other is the isospin singlet $dg$-$sb$-$ur$ mode.

Next qualitative change occurs when $T$ is increased to $T\sim21.2\MeV$.
At this point, the $ub$-$sr$ mode with $\Qtilde=+1$ becomes gapless
and the guSC(5) phase sets in.
Notice that because the thermally-excited quasi-quarks are
already present in the system over the range
($E_{ub\mbox{\scriptsize-}sr}\alt T$), there is no sharp boundary
between the uSC(6) and guSC(5) phases.

At a little higher temperature $T\sim 21.8\MeV$, $\De_2$ vanishes and
above which the 2SC(4) phase is realized. 
At this transition point, the $sr$-$ub$ mode having $\Qtilde=+1$ becomes
gapless.
Through the gap parameter $\De_2$, this $sr$-$ub$ mode is paired with
its partner, \ie, the gapless $ub$-$sr$ mode.
When $\De_2\to 0$, these two modes get unpaired to become the bare $ub$
and $sr$ quarks.
This guSC(5) $\to$ 2SC(4) phase transition is of second order.

As $T$ is increased further, the crossover 2SC(4) $\to$ g2SC(2) takes
place at $T\sim26.6\MeV$, and finally the g2SC(2) is
taken over by the UQM(0) phase through a second order phase transition 
at $T\sim27.4\MeV$.

We have found that the excitation gaps behave in a more complicated
manner than the gap parameters as functions of $T$.
However, as we explained, there are no sharp boundary with
thermodynamical singularity between CFL(9) and gCFL${}_8$(8), uSC(6) and
guSC(5), and 2SC(4) and 2SC(2) transitions.
For this reason, we did not indicate these crossover boundaries in the
phase diagram of FIG.~\ref{phase_weak}.

\begin{figure}[thp]
  \includegraphics[width=0.42\textwidth,clip]{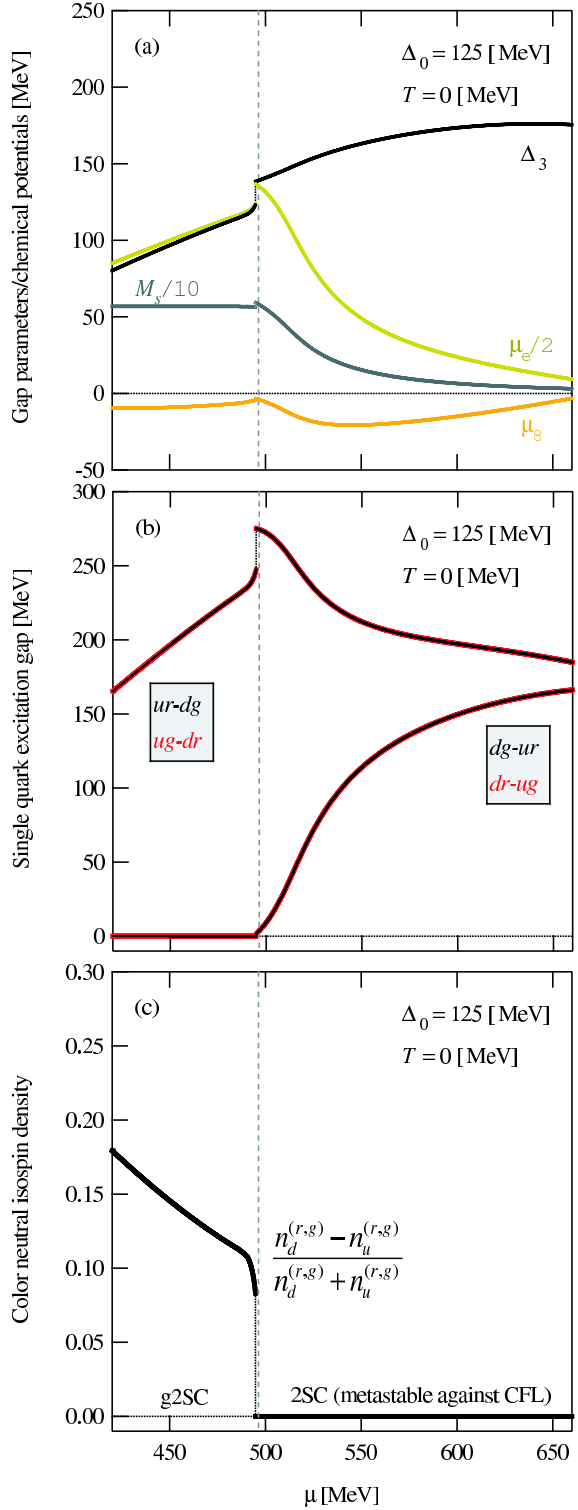}
 \caption[]{{\bf (a)}~~The gap and mass parameters, and the chemical
 potentials versus $\mu$ along the (g)2SC solution. 
 The dashed line placed on $\mu=496.4\MeV$ represents the point 
 above which the 2SC state is metastable against the CFL.
 {\bf (b)}~~The four quasi-quark gaps as a function of $\mu$. 
 {\bf (c)}~~The fraction of the isospin density
 $\sum_{a=r,g}(n_d^{a}-n_u^{a})$ to the iso-scaler
 density $\sum_{a=r,g}(n_u^{a}+n_d^{a})$; by the summation, this
 quantity is made of neutral with respect to the remaining
 SU$(2)_{\mbox{\scriptsize color}}$ charges.
 }
 \label{gapvsmu1}
\end{figure}
\begin{figure}[htp]
  \includegraphics[width=0.42\textwidth,clip]{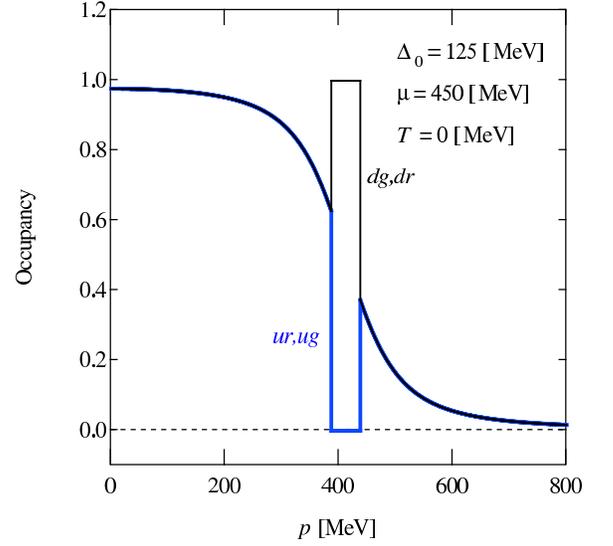}
 \caption[]{The occupation number for $u$ and $d$ quarks in the 
 pairing $(r,g)$ sector as a function
 of momentum $p$ in the g2SC phase 
 at $\De_0=125\MeV$, $\mu=450\MeV$ and $T=0$. 
 }
 \label{occup}
\end{figure}
\begin{figure}[htp]
  \includegraphics[width=0.42\textwidth,clip]{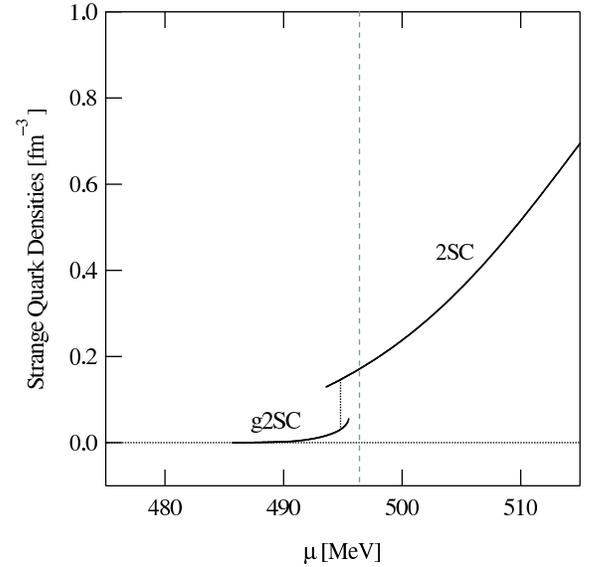}
 \caption[]{%
 The strange quark density as a function of $\mu$
 in the 2SC sector. The vertical dashed line is the same as that in
 FIG.~\ref{gapvsmu1}. At $\mu=498.4\MeV$, the strange quark density
 jumps accompanying the first order 2SC/g2SC transition.
 }
 \label{SQDensity}
\end{figure}
\begin{figure}[htp]
  \includegraphics[width=0.42\textwidth,clip]{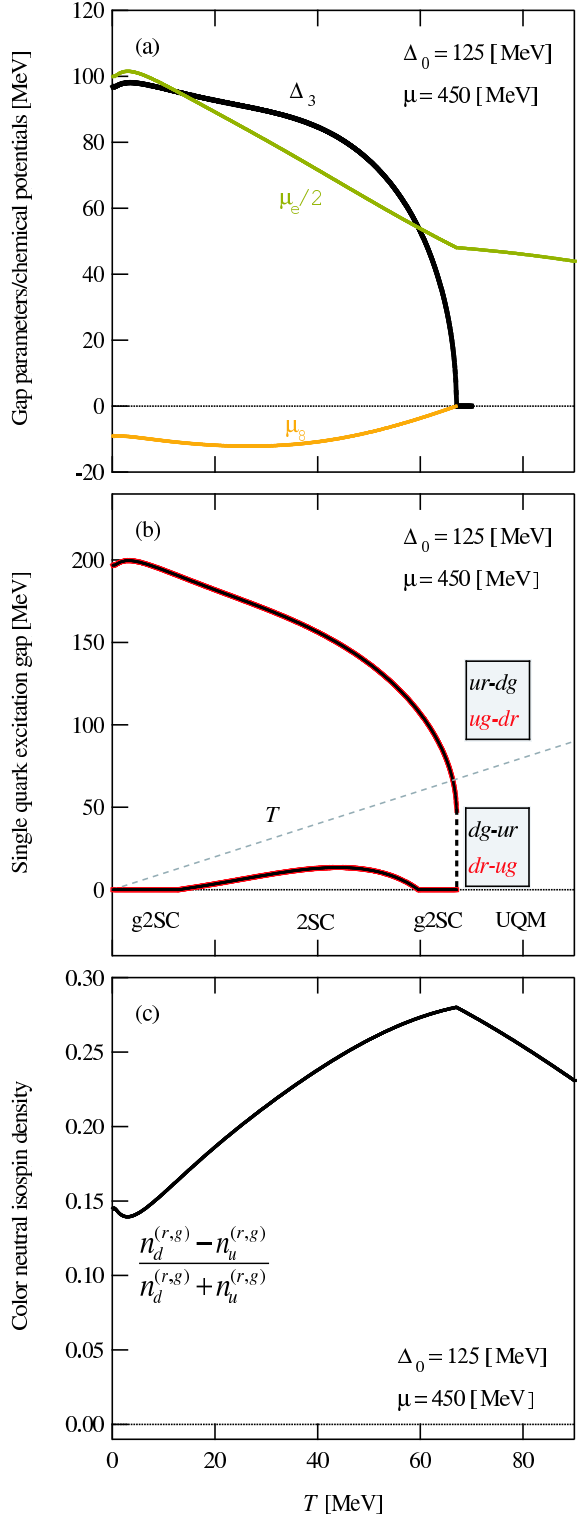}
 \caption[]{%
 {\bf (a)}~~The gap parameter $\De_3$ and the chemical potential
 $\mu_{e,8}$ versus $T$ for $\De_0=125\MeV$ and $\mu=450\MeV$.
 {\bf (b)}~~The excitation gaps in the four quasi-quark spectra as 
 a function of $T$.
 {\bf (c)}~~The SU$(2)_{\mbox{\scriptsize color}}$ neutral 
 isospin density divided by the total isoscalar density 
 as a function of $T$. 
 This quantity is an order parameter for the 2SC/g2SC transition at
 $T=0$.
 }
 \label{gapvsT0}
\end{figure}
\begin{figure}[htp]
  \includegraphics[width=0.42\textwidth,clip]{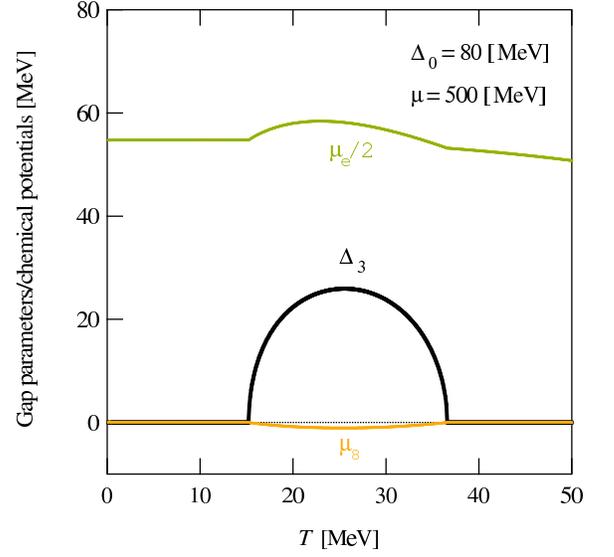}
 \caption[]{%
 The gap parameter $\De_3$ and the chemical potentials 
 $(\mue/2,\mu_8)$ versus temperature $T$ 
 at $\mu=500\MeV$ and $\De_0=80\MeV$; weak coupling case.
 }
 \label{Gap_D080mu500}
\end{figure}
\begin{figure}[htp]
  \includegraphics[width=0.42\textwidth,clip]{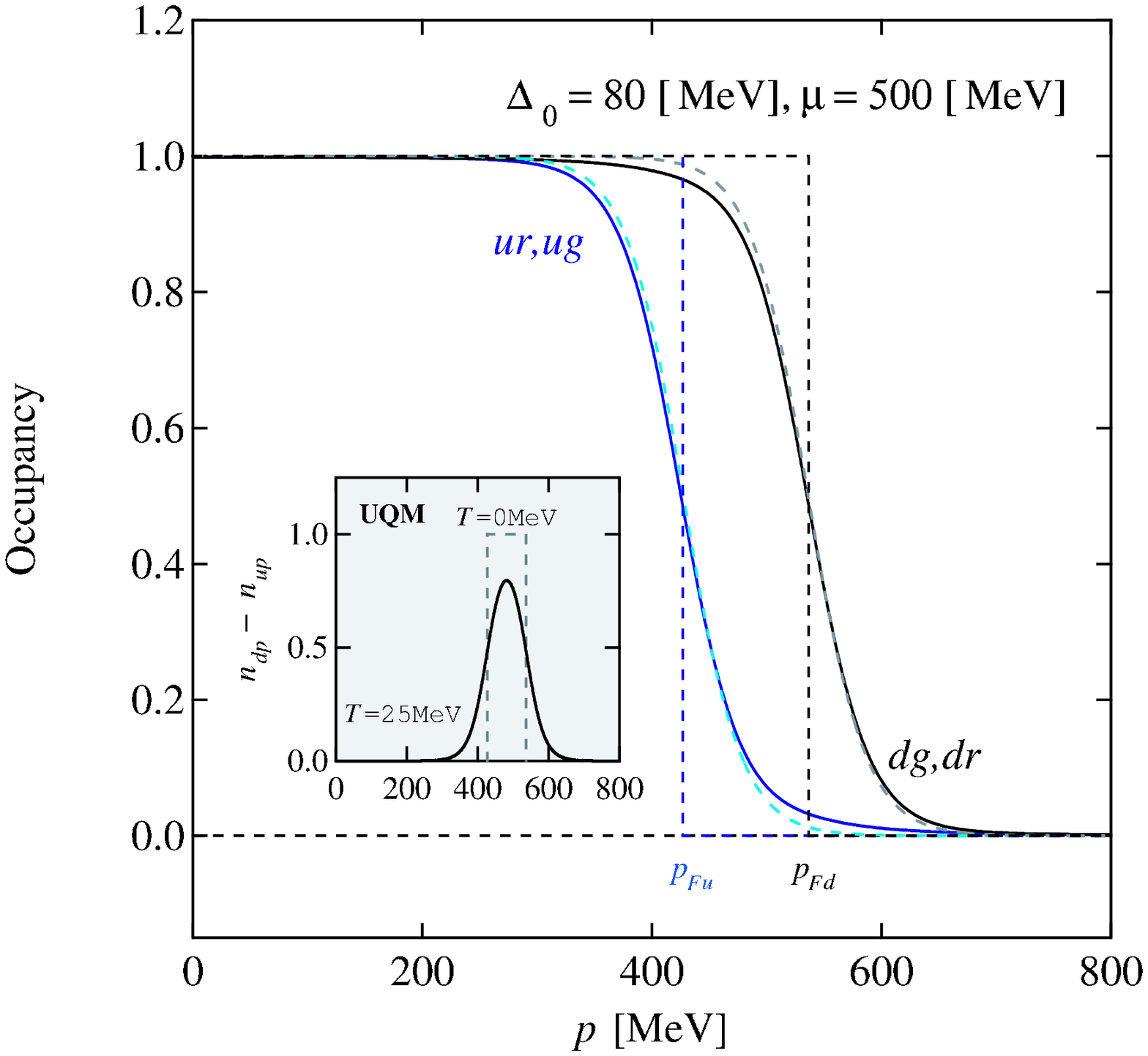}
 \caption[]{%
 Solid lines are the occupation numbers for the $u$ and $d$ quarks
 in the g2SC phase $(\mu,T)=(500\MeV,25\MeV)$ as a function of 
 momentum $p$.
 The sharp Fermi surfaces for the $u$ and $d$ quarks at $T=0$ 
 are also indicated by the dashed lines.
 Dashed curves are the $u$ and $d$ occupancies for the {\em unstable}
 UQM phase at $T=25\MeV$. 
 The inset is the quantity $n_{dp}^{(r,g)}-n_{dp}^{(r,g)}$, \ie, the
 isospin mismatch in the UQM at $T=0$ (dashed line) and for the
 unstable UQM at $T=25\MeV$ (solid line).
 }
 \label{Occup_D080T25}
\end{figure}

\subsection{Phases for weak and intermediate coupling}\label{weakint}
Let us next examine the weak and intermediate coupling cases ($\De_0=80$
and $\De_0=125\MeV$).
In FIG.~\ref{phase_int}, the phase diagrams for the both cases are
shown.
\subsubsection{Phases for the intermediate coupling}\label{intcoupling}
We first discuss  the case of the 
intermediate coupling $\De_0=125\MeV$, in advance of the weak coupling
case. 

FIG.~\ref{phase_int}(b) shows that the UQM and gCFL phases disappear
at $T=0$ and the g2SC (CFL) phase exist in the intermediate (high)
density regime \cite{Abuki:2004zk}.
We notice that the fully gapped 2SC phase exists in a small $\mu$-region
between the CFL and g2SC phases. 
Also we notice that the transition between the 2SC and g2SC phases is of
a first order accompanied by a jump in the dynamical strange quark
mass and other physical quantities.
This is in contrast to the usual 2SC/g2SC transition without strange
quarks.
We will later discuss this point in detail.

The phases for $T\ne 0$ also differ from the extremely weak coupling
case: (i) The window for the uSC phase is pushed away to higher
density side and is confined in a small region.
The cross placed on the dashed line represents the point at which
the window for the uSC phase opens.
(ii) The large dot putted on the CFL/2SC transition line indicates 
the critical endpoint; the dashed line between the large dot and the
cross shows that there exists the continuous phase transition of
2SC $\leftrightarrow$ CFL.
In the following, we shall discuss the detail of the 2SC/g2SC transitions.
We first study the $T=0$ case and then investigate the $T\ne0$ case.

\setcounter{paragraph}{0}
\vspace*{0.3cm}
\paragraph{2SC/g2SC transition at $T=0$:}
In FIG.~\ref{gapvsmu1}(a), we explicitly show the solution
of the gap equation in the 2SC sector at $T=0$.
Just to make the argument clear, we show the 2SC solution for the entire
$\mu$-space, but note that the 2SC phase is metastable being
energetically taken over by the CFL phase in the right hand side of the
vertical dashed line placed on $\mu=496.4\MeV$; we should keep in mind
that the window for the 2SC phase is extremely narrow.
Below the critical chemical potential $\mu=494.8\MeV$,
$(\mu_{ug}-\mu_{dr})/2=(\mu_{ur}-\mu_{dg})/2=\mue/2$ becomes greater
than the gap parameter $\De_3$, and as a consequence, the fully gapped 2SC is
taken over by the partially ungapped g2SC (See FIG.~\ref{gapvsmu1}(b)).
In FIG.~\ref{gapvsmu1}(c), the ratio of the SU$(2)_{\mbox{\scriptsize
color}}$ neutral isospin density $\sum_{a=r,g}(n_d^a-n_u^a)$ relative to
the isospin scaler density is shown as a function of $\mu$.
This quantity serves as an order parameter with which the g2SC phase can
be distinguished from the 2SC phase;
in the fully gapped 2SC phase, this quantity must be zero because the
equal number of $u$ and $d$ quarks are required for the pairing, while
in the g2SC phase, the quantity becomes nonzero because of the
accumulation of isospin charge in the blocking region.
This is clearly seen in the plot of the occupation numbers in the g2SC
phase (See FIG.~\ref{occup}).
From all the three figures (FIG.~\ref{gapvsmu1}) in addition to the
analyses of the thermodynamic potential, we can conclude that the
2SC/g2SC transition is of first order unlike the usual case without 
the chiral dynamics, \ie, the $\lt\bar{s}s\gt$ condensation.

The first order phase transition between these two phases is brought
about actually by the competition between the chiral $\lt\bar{s}s\gt$
condensation and the quark pairing dynamics in the $u$-$d$ sector
as was first recognized in \cite{Abuki:2004zk}.
We can see a small jump in the strange quark mass at the transition
point in FIG.~\ref{gapvsmu1}(a).
The strange quark mass reaches almost the vacuum value
on the g2SC side, while it varies on the 2SC side and becomes 
smaller as the density goes high.
Accordingly the strange quarks are absent in the g2SC side, while 
a small number of $s$ quarks are present in the 2SC side and it 
grows with increasing density.
This is naturally understood as follows:
In the fully gapped 2SC phase, the $u(rg)$ and $d(rg)$ quark
densities should be equal for accommodating the $\De_3$-condensate.
Because of the positive electric charge $1/3=2/3-1/3$ coming from
this $SU(2)_{\mbox{\scriptsize color}}$ sector, the system
needs strange quarks, electrons and $db$-quarks.
Thus, the high density condition [$\Ms\ll\mu$] favors the 2SC
realization because plenty of strange quarks providing negative charges
can exist in the system as can be confirmed by the plot of strange
quark density in FIG.~\ref{SQDensity}. 
But this condition is in turn disfavored with respect to 
the chiral condensation energy.
On the other hand, the g2SC phase can realize the electric neutrality
with a less number of strange quarks because 
of a $d$ quark excess to $u$ quarks
in the $SU(2)_{\mbox{\scriptsize color}}$ sector.
Accordingly the g2SC phase is not so much disfavored by the chiral
$\lt\bar{s}s\gt$ condensation ($\Ms\agt\mu$) in comparison with the 
2SC phase \footnote{%
One may think it somewhat surprising that the strange
quark mass jumps downwards going from 2SC to g2SC, which seemingly 
contradicts the discussion that plenty of strange quarks prefer the 2SC phase. 
However this is not the case as can be seen from FIG.~\ref{SQDensity}. 
In fact, the Fermi momentum squared for $sb$ quark is given
by $p_{Fsb}^2=(\mu+\mue/3-2\mu_8/3)^2-\Ms^2$. 
When one goes from the 2SC side to the g2SC side, $\Ms$ certainly
drops a little, but $\mue(\sim 2\De_3)$ also drops at the transition point.
The latter effect on the $p_{Fsb}$ wins over the former effect
so that the strange quark density jumps downwards.
Accordingly, the dynamical chiral condensation energy due to
$\lt\bar{s}s\gt$ gets slightly lost in the g2SC side. 
This energy loss is compensated by the reduction of the kinematical
energy cost due to the stress arising from pinning the $u$- and $d$-Fermi
momenta in the $SU(2)_{\rm\scriptsize color}$ sector at equal level in
the 2SC side (see the decrease of $\mue$ from 2SC to g2SC). 
As a consequence of these competing dynamical and kinematical ingredients, 
$\Om_{\rm 2SC}=\Om_{\rm g2SC}$ holds at
$\mu=494.8\MeV$, \ie, the 2SC/g2SC transition point.}.
In short, the g2SC phase can co-exist with the chiral symmetry broken
phase easier than the fully gapped 2SC phase can do.
The competition of these two ingredients makes the 2SC/g2SC transition
be of first order at $T=0$.

\vspace*{0.3cm}
\paragraph{g2SC/2SC/g2SC crossovers for $T\ne0$:}
We now investigate how the situation changes for the 2SC/g2SC 
transition at $T\ne0$.
The phase diagram displayed in FIG.~\ref{phase_int}(b) shows that, in a
relatively low density regime ($430\MeV\alt\mu\alt480\MeV$), the
g2SC/2SC and 2SC/g2SC transitions take place successively when the quark
matter is heated.
This kind of exotic situation is already found in \cite{Shovkovy:2003uu}
where the systematic study of the coupling strength dependence of the
2SC/g2SC transition is done within the two-flavor NJL model.
In our case, the strange quark mass reaches almost its vacuum value
in this region so that the pairing dynamics becomes similar to those
in the previous two-flavor models.
In FIG.~\ref{gapvsT0}, we show how the physical quantities behave 
through these phase transitions for $\mu=450\MeV$. 
FIG.~\ref{gapvsT0}(a) shows the gap parameters and the chemical
potentials as a function of $T$. 
$\mue/2$ and $\De_3$ cross each other twice with increasing $T$ so that
the system undergoes the 2SC/g2SC and g2SC/2SC transitions before
the pairing is overwhelmed by the UQM.
FIG.~\ref{gapvsT0}(b) shows the gaps in the quasi-quark spectra.
We note, however, that the excitation gap in the $dg$-$ur$ ($dr$-$ug$)
quasi-quark spectrum is always smaller than $T$ so that these
quasi-quarks are thermally-excited irrespective of whether the system is
in the 2SC phase or the g2SC phase.
These quasi-quarks smear the 2SC/g2SC and g2SC/2SC transitions.
Accordingly, no thermodynamic singularity is associated with these
transitions except for the final g2SC-UQM ($\De_3\to0$) transition.
This can be understood also in terms of the order parameter for
the 2SC/g2SC transition at $T=0$.
FIG.~\ref{gapvsT0}(c) shows the SU$(2)_{\mbox{\scriptsize color}}$
neutral isospin density $\sum_{a=r,g}(n_d^a-n_u^a)$ normalized 
by isoscalar density as a function of $T$.
From the figure, we can see that the g2SC/2SC transitions are 
crossovers; the order parameter always takes nonzero value
and the system is always isospin charged due to the 
thermally-excited quasi-quarks even for $\mue/2<\De_3$ region.

\subsubsection{Phases for the weak coupling}\label{weakcoupling}
Now we turn to the weak coupling case ($\De_0=80\MeV$).
The phase diagram is shown in FIG.~\ref{phase_int}(a).
In this case, the diquark coupling is not stronger enough to exclude the
UQM phase from $T=0$, 
while the gCFL phase is overwhelmed by the UQM with a large
$\lt\bar{s}s\gt$ and thus is washed out as was first noted in the
previous work \cite{Abuki:2004zk}.

One of the most striking features in this case is the appearance of 
the g2SC phase for $T\ne0$ even when the UQM phase is realized at $T=0$.
This somewhat strange aspect of the color-neutral g2SC phase was
first recognized in \cite{Shovkovy:2003uu} and has been further
confirmed in \cite{Fukushima:2004zq}, although it seems that physics 
for this phenomenon has not yet been clarified enough.
In FIG.~\ref{Gap_D080mu500}, we show the gap parameter $\De_3$
and the chemical potential $(\mue/2,\mu_8)$ as a function of 
$T$ in this situation ($\mu=500\MeV$).
At $T=0$, the system is in the UQM phase, and the $\De_3$-condensate
appears at $T\sim16\MeV$ and then the 2SC turns again into the UQM
phase at $T\sim36.5\MeV$.
$\mue/2$ is always larger than $\De_3$ so that the system is
in the ``g2SC'' phase accompanied by only two gapped quasi-quarks,
\ie, the $ur$-$dg$ and $ug$-$dr$ modes.
However, it should be again noted here that the two phases (2SC and
g2SC) are not thermodynamically distinguishable for $T\ne0$ as 
explained in the previous section.
To understand the appearance of the 2SC phase for $T\ne0$, we show
the occupation number of the $u$ and $d$ quarks for various situations
in FIG.~\ref{Occup_D080T25}.
There are sharp Fermi surfaces at $T=0$ which are indicated by dashed
lines, \ie, one for $d$ quarks at $p_{Fd}=\mu+\mue^{(T=0)}/3$ and the
other for $u$ quarks at $p_{Fu}=\mu-2\mue^{(T=0)}/3$.
The dashed curves show the $u$ and $d$ quark occupancies at $T=25\MeV$
in the UQM phase which is unstable to the formation of $u$-$d$ pairs.
Quarks are thermally-excited so that the distribution is somewhat
smeared.
We note that because of this thermal effect, the $u$ and $d$ quarks
are under a better kinematical matching than the situation at $T=0$;
from the inset of the figure, we can see that the isospin mismatch
$(u_d-u_u)$ at the averaged Fermi momentum $p=(p_{Fd}+p_{Fu})/2$
is reduced in the $T=25\MeV$ case from the unity, \ie, the value
in the $T=0$ case.
As a result, it is easier  for the quark matter in the
UQM phase to form the diquark condensate in the
isospin singlet channel at finite temperature than  at $T=0$.
The solid lines show the occupation numbers after this reconfiguration
takes place (the g2SC phase).
We remark that unlike the occupation numbers in the g2SC phase at $T=0$
drawn in FIG.~\ref{occup}, there are no singular points in those
for the thermally-smeared g2SC phase for $T\ne0$.

\subsection{Phases for (extremely) strong coupling}\label{stcoupling}
\begin{figure}[thp]
  \includegraphics[width=0.42\textwidth,clip]{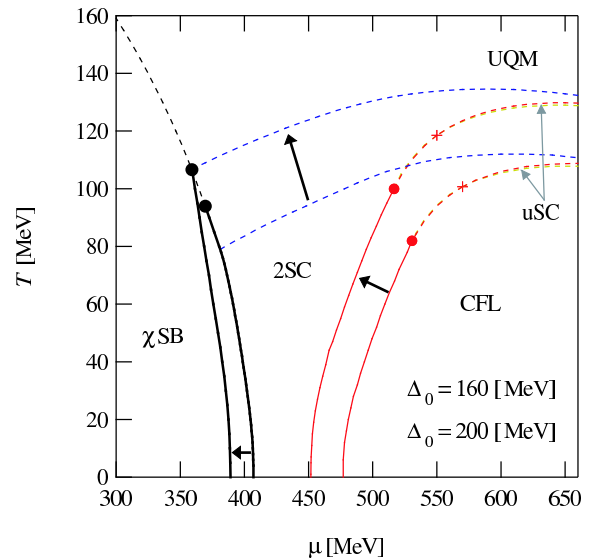}
  \caption[]{The phase diagrams for the strong coupling
  $(\De_0=160\MeV)$ and for the extremely strong coupling
  $(\De_0=200\MeV)$. 
  Phase boundaries shift in the directions indicated by the bold arrows
  when the diquark coupling is increased from $\De_0=160\MeV$ to
  $\De_0=200\MeV$.}
  \label{fig:02}
\end{figure}
We next discuss the strong and extremely strong coupling cases,
$\De_0=160\MeV$ and $200\MeV$, respectively.
The phase diagrams are depicted in FIG.~\ref{fig:02}.
The phase boundaries move in the directions indicated by the bold arrows
when the diquark coupling is increased from $\De_0=160\MeV$ to
$\De_0=200\MeV$. 
One can see that both the phase structures in these cases
are qualitatively identical and are much simpler than those 
for the weak and intermediate couplings.
The premature gapless phases disappear from the low temperature regime,
and only the two major pairing phases remain in the phase diagram,
\ie, the CFL phase and the 2SC phase.
In each case, we still have a small window for the uSC realization
as a precursory phase of the CFL $\to$ 2SC transition in the high
$\mu$-region of the phase diagram.
However, one should notice that 
 the critical temperature 
for the 2SC/UQM transition has a maximum in this chemical potential region,
which might imply that this shrinkage of the uSC window is due to 
a  cutoff artifact.
It should be also noted that when the diquark coupling is large, the 2SC
phase encroaches upon the domain for the $\chi$SB phase.
In addition, the 2SC/UQM transition temperature becomes larger with
increasing diquark coupling, and at some critical coupling
($\De_0=\De_{\rm c}$), it reaches
the tricritical point of the chiral phase transition:
At this coupling, the chiral tricritical point acquires the nature of
the doubly critical point where the two second order critical
lines merge; one is for the 2SC/UQM transition and the other is 
for the UQM/$\chi$SB transition.
Because this point has both the tricritical and doubly critical 
natures, we call it the ``TDCP'' point.
When the diquark coupling is increased beyond this critical
coupling $\De_{\rm c}$, the TDCP point shifts to higher temperature
because the 2SC/$\chi$SB transition is always of first order
\cite{Ruster:2005jc}.
Our calculation indicates the critical coupling $\De_{\rm c}$ where
the tricritical point obtains a DCP nature lies between $\De_0=160\MeV$
and $\De_0=200\MeV$.

Finally, it is worth mentioning that  
the phases in the strong coupling cases discussed above
are all free from the instability problem associated with 
the imaginary Meissner masses \footnote{%
It is easily verified that there is no region where the 
instability condition at $T=0$, i.e.,
($\sqrt{2}\De_{\rm 2SC}(\mu)<\mue$) \cite{Unstable},
holds.
As for the ($T\ne0$) case, we need a 
numerical evaluation of the Meissner Masses to verify the stability since 
there is no such analytic formula as above for the instability condition.
The numerical calculations show, however,
that the singularity associated with the gapless modes is
smoothed out by thermal quasi-particles at $T\ne0$, which
implies that the system at $T\ne0$ is free from the
instability as long as that at $T=0$ is stable
\cite{Unstable,Fukushima:2005cm}.
Thus one can conclude that there is no instability in the entire phase
diagram for the (extremely) strong coupling case.
The strategy adopted in \cite{Sandin:2005um} might also be useful to
specify the instability condition.
}.

\subsection{Baryon density $\rho_B$ at $T=0$}\label{quarknumber}
\begin{figure}[htp]
  \includegraphics[width=0.42\textwidth,clip]{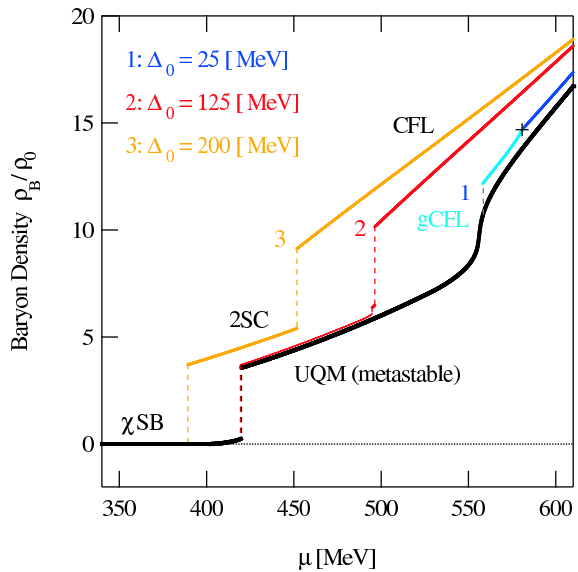}
  \caption[]{The baryon number density at $T=0$ in the unit of
  $\rho_0=0.17\,{\rm fm}^{-3}$, \ie, the normal nuclear density.
  $1$, $2$, and $3$ indicated for each line correspond to the
  following three cases; the extremely weak coupling ($\De_0=25\MeV$),
  the intermediate coupling ($\De_0=125\MeV$), and the extremely strong
  coupling ($\De_0=200\MeV$), respectively. 
  The cross placed on the CFL line indicates the transition point at
  which the CFL continuously turns into the gCFL phase.
  }
  \label{quarkdensity}
\end{figure}
We have investigated the phase structures of the quark matter 
in the $(\mu,T)$-plane for several diquark couplings.
It is, however, sometimes more convenient to describe the system
in the $(\rho_B, T)$-plane to have a physics intuition into
the system.
Also it is necessary to have the equation of state
as a function of $\rho_B$ to clarify 
the inner structure of compact stars \cite{Blaschke:2005uj}.
Here one should notice that when multiple phases coexist at a temperature
with the same chemical potentials, each phase may have a 
different baryon density from each other; these phases are actually
realized in a mixed phase where the phases with different baryon
densities coexist.

Here, we ask how dense each phase at $T=0$ is.
In FIG.~\ref{quarkdensity}, we show the baryon density $\rho_B$ 
versus chemical potential $\mu$ at $T=0$ in the following three cases;
(1) the extremely weak coupling $\De_0=25\MeV$, 
(2) the intermediate coupling $\De_0=125\MeV$ and 
(3) the extremely strong coupling $\De_0=200\MeV$.
In any case, as the chemical potential is lowered from the
highest value, the baryon density decreases and shows
several jumps associated with first order transitions.
We start with the case (1). 
In this case, the ground state at highest density, about 15 times the
normal nuclear matter density, is in the CFL phase.
When the density is decreased down to the point indicated by the
cross, the CFL continuously turns into the gCFL phase with a 
steeper gradient
$\frac{d\rho}{d\mu}=\frac{1}{\rho}\big(\frac{d\rho}{dP}\big)_{%
\scr T}\equiv\ka_{\scr T}(\rho)$, \ie, a larger
compressibility or equivalently, a larger number fluctuation
(susceptibility $\chi_{\rho\rho}^T(r)$) attributed to gapless modes. 
If the density is decreased further, the first order transition
from the gCFL to the UQM takes place; accordingly the baryon density 
drops down to almost ten times the nuclear density through a mixed 
phase of the two phases.
The UQM phase continues down to $\sim 3$ times the nuclear density,
and then gets taken over by the chiral symmetry broken phase.
The situation becomes slightly complicated in the case (2).
The densest phase is again the CFL.
As the density is decreased, the CFL phase persists down to
almost ten times the nuclear density. 
Then the quark system undergoes a first order phase transition
to the 2SC phase via the dynamical nucleation and growth of the low
density 2SC droplets in the CFL phase.
This nucleation also needs a lot of weak processes.
The density window accommodating the 2SC phase is extremely small
and the 2SC phase gets immediately taken over by the g2SC phase.
The g2SC phase finally changes into the $\chi$SB phase.
Similarly in the case (3), the baryon density drops twice according to
the CFL/2SC and 2SC/$\chi$SB transitions. 
It is interesting that the lowest baryon density of the CFL phase
decreases with increasing diquark coupling while the density at the same
$\mu$ is enhanced by the factor $\De_0^2/\mu^2$.
It should be noted, however, that the $\rho$-$\mu$ relation will
be affected by the fluctuation effect so that the density becomes
even larger due to bosonic degrees of freedom in the strong
coupling cases \cite{Nishida:2005ds}.

\subsection{The Ginzburg-Landau analysis}\label{GLtheory}
In this section, we make a systematic Ginzburg-Landau analysis in order
to understand why the dSC phase does not appear in our calculation of
the phase diagram.
Our analysis is an extension of the previous work \cite{Iida:2003cc};
we expand the Ginzburg-Landau coefficients up to quartic order in $\Ms$
which were not taken into account in \cite{Iida:2003cc} but turn out to
play an important role for the relatively strong coupling (low density)
regime.
We also notice that this extension makes it possible to provide a
unified and systematic description of the thermal pairing/unpairing
transitions obtained in the NJL \cite{Ruster:2005jc} and the
Ginzburg-Landau analysis \cite{Iida:2003cc}.

If the phase  transition at finite temperature is of second order 
where the condensate $\De_\eta$ vanishes at the critical point,
we can expand the effective potential in terms of the gap parameters
near the critical temperature $T_c$.
In this section, we focus on the regime where the chiral symmetry is
restored, and assume that the thermal pairing/unpairing transitions 
are not so much affected by the chiral dynamics there.
Thus, we shall switch off the scalar coupling $G_s$ in our NJL
lagrangian, \Eqn{eq:lag}, and treat the strange quark mass $\Ms$
as a constant parameter. 
This is nothing but a model used in
\cite{Fukushima:2004zq,Fukushima:2005fh}, which will be referred as the
{\em diquark} NJL model hereafter in order to distinguish it from the
present NJL model analysis with the scalar coupling, \ie,
\Eqn{eq:lag}.
Treating $\Ms$ as a parameter can be justified by the fact that $\Ms$
does not so much vary in the vicinity of $T_{c}$ (See
FIG.~\ref{gapT_weak}(a)).
Either from the diquark NJL model or from the QCD-like theory
with the Cornwall-Jackiw-Tomboulis potential up to 2PI graphs
\cite{Iida:2003cc}, we can derive the Ginzburg-Landau potential.
We first emphasize that each of the Ginzburg-Landau coefficients 
becomes a function of $\Ms$ under the neutrality constraints.
We evaluate these functions by the Taylor expansion in $\Ms$.
Our final task is to calculate the splittings in the critical
temperatures up to quartic order in $\Ms$.
Here, we shall only give our result for the Ginzburg-Landau potential
leaving the detail of the calculation to the Appendix \ref{GLapproach},
because it is somewhat involved although straightforward.
After solving the neutrality constraints, the Ginzburg-Landau potential
is found to be
\begin{widetext}
\bea
 {\mathcal L}_{\rm GL}&=&4N[\mu]\De_1^2\left(\frac{T-T_{c0}}{T_{c0}}+\frac{\Ms^2}{6\mu^2}%
  \log\left(\frac{\mu}{T_{c0}}\right)+%
  \frac{7\ze(3)\Ms^4}{64\pi^2\mu^2T_{c0}^2}\right)\nn
  &&+4N[\mu]\De_2^2\left(\frac{T-T_{c0}}{T_{c0}}+\frac{7\Ms^2}{24\mu^2}%
  \log\left(\frac{\mu}{T_{c0}}\right)+%
  \frac{7\ze(3)\Ms^4}{256\pi^2\mu^2T_{c0}^2}\right)\nn
  &&+4N[\mu]\De_3^2\left(\frac{T-T_{c0}}{T_{c0}}+\frac{\Ms^2}{24\mu^2}%
  \log\left(\frac{\mu}{T_{c0}}\right)+%
  \frac{7\ze(3)\Ms^4}{256\pi^2\mu^2T_{c0}^2}\right)\nn
  &&+\frac{7\ze(3)N[\mu]}{8\pi^2T_{c0}^2}%
      \left(\De_1^4+\De_2^4+\De_3^4\right)%
  +\frac{7\ze(3)N[\mu]}{8\pi^2T_{c0}^2}%
      \left(\De_1^2+\De_2^2+\De_3^2\right)^2,
\label{GLEXP}
\eea
\end{widetext}
where $N[\mu]=\mu^2/2\pi^2$ is the density of state and $T_{c0}$ is the
critical temperature for the symmetric ($\Ms=0$) quark matter. 
Some remarks are in order here. 
(i) The coefficients for the $\De^4$ terms are not expanded in $\Ms$.
This is because those effects on the splittings of $T_{c0}$ are
small with a suppression factor $(T_{c0}/\mu)^4$
in comparison with the contribution from $\Ms$-dependent terms 
in the coefficients of $\De^2$ terms.
(ii) Up to quadratic order in $\Ms$, this exactly coincides with the
form obtained in the previous study \cite{Iida:2003cc} as it should be
\footnote{%
It is not strange that we have arrived at the same expression for the
Ginzburg-Landau potential as that in \cite{Iida:2003cc} although 
we have derived it from the diquark NJL model with the contact
two fermion interaction.
This is simply because the non-locality of the gluon-mediated attraction
is incorporated into the Ginzburg-Landau potential only through
$T_{c0}$ and the gap function $\De_0(p)$ with $p$ being the momentum of
quasi-quarks in \cite{Iida:2003cc}.}.
(iii) The $\Ms^4$-correction to the Ginzburg-Landau coefficient is not
 so small because it is not a simple expansion in
 $\Ms/\mu$; the coefficient of $\Ms^4$ is enhanced by factor
 $\mu^2/T_{c0}^2$.

We now calculate the splittings of critical temperature $T_{c0}\to
(T_{c1},\,T_{c2},\,T_{c3})$. 
We first solve the gap equation
\bea
  \frac{\p {\mathcal L}_{\rm GL}}{\p \De_1^2}=0,\quad%
  \frac{\p {\mathcal L}_{\rm GL}}{\p \De_2^2}=0,\quad%
  \frac{\p {\mathcal L}_{\rm GL}}{\p \De_3^2}=0,\quad%
\eea
in $\De_1^2,\De_2^2,\De_3^2$ and define the temperatures
$(T_{\rm ds},T_{\rm su},T_{\rm ud})$ as the  solutions of 
$\De_1^2(T_{\rm ds})=0,\De_2^2(T_{\rm su})=0,\De_3^2(T_{\rm ud})=0$.
Up to the quartic order in $\Ms$, we obtain
\beq
\ba{rcl}
  T_{\rm ds}&=&T_{c0}\left[1-\log\left(\frac{\mu}{T_{c0}}\right)%
               \frac{\Ms^2}{6\mu^2}-%
               \frac{35\ze(3)\mu^2}{128\pi^2T_{c0}^2}%
               \frac{\Ms^4}{\mu^4}\right],\\[2ex]
  T_{\rm su}&=&T_{c0}\left[1-\log\left(\frac{\mu}{T_{c0}}\right)%
               \frac{2\Ms^2}{3\mu^2}+%
               \frac{7\ze(3)\mu^2}{128\pi^2T_{c0}^2}%
               \frac{\Ms^4}{\mu^4}\right],\\[2ex]
  T_{\rm ud}&=&T_{c0}\left[1+\log\left(\frac{\mu}{T_{c0}}\right)%
               \frac{\Ms^2}{3\mu^2}+%
               \frac{7\ze(3)\mu^2}{128\pi^2T_{c0}^2}%
               \frac{\Ms^4}{\mu^4}\right].\\[1ex]
\ea
\eeq
Here we can also see that the $\Ms^4$-correction could give a
significant contribution comparable to the $\Ms^2$ term 
because of an enhancement factor $\mu^2/T_{c0}^2\log(\mu/T_{c0})$.
We can see that $T_{\rm ud}$ is largest of the three temperatures, which
simply means that the 2SC phase is robustest against the thermal
disturbance.
We can evaluate the critical temperature $T_{c3}$ for the 2SC $\to$ UQM
transition by putting $\De_1=\De_2\to0$ in \Eqn{GLEXP} and solving
$\frac{\p{\mathcal L}}{\p{\De_3^2}}=0$ in $T$.
The result is
\beq
   T_{c3}=T_{c0}\left[1-\log\left(\frac{\mu}{T_{c0}}\right)%
               \frac{\Ms^2}{24\mu^2}-%
               \frac{7\ze(3)\mu^2}{256\pi^2T_{c0}^2}%
               \frac{\Ms^4}{\mu^4}\right].
\eeq
In contrast to the fact that the 2SC is always the hottest pairing
phase, the second pairing phase next to the 2SC depends on the
value of the strange quark mass.
We have to consider the following two cases;
\beq
\ba{lll}
  \mbox{case [A]~:~} & 
            \dsp\Ms^2<\frac{32\pi^2}{21\ze(3)}%
                    T_{c0}^2\log\left(\mu\big/T_{c0}\right) &%
		    \to\,T_{\rm su}<T_{\rm ds}.\\[2ex]
  \mbox{case [B]~:~} & 
            \dsp\Ms^2>\frac{32\pi^2}{21\ze(3)}%
                    T_{c0}^2\log\left(\mu\big/T_{c0}\right) &%
		    \to\,T_{\rm ds}<T_{\rm su}.\\[1ex]
\ea
\nonumber
\eeq
In case [A], $\De_2$ first becomes zero at $T_{\rm su}\equiv T_{c2}$
and then $\De_1$ vanishes at $T_{c1}(>T_{c2})$ when the quark matter
is heated from the CFL phase; that is
\bea
\ba{rcl}
  T_{c2}&=&\dsp T_{c0}\left[1-\log\left(\frac{\mu}{T_{c0}}\right)%
               \frac{2\Ms^2}{3\mu^2}+%
               \frac{7\ze(3)\mu^2}{128\pi^2T_{c0}^2}%
               \frac{\Ms^4}{\mu^4}\right],\\[2ex]
  T_{c1}&=&\dsp T_{c0}\left[1-\log\left(\frac{\mu}{T_{c0}}\right)%
               \frac{7\Ms^2}{24\mu^2}-%
               \frac{49\ze(3)\mu^2}{256\pi^2T_{c0}^2}%
               \frac{\Ms^4}{\mu^4}\right].\\[2ex]
\ea
\eea
This is actually the case which is studied in \cite{Iida:2003cc};
the quark matter undergoes a hierarchical unlocking CFL $\to$ dSC
$\to$ 2SC $\to$ UQM.
We note, however, that the quartic terms are derived for the first time
in this study and these will turn out to be a crucial for a 
unified picture of the thermal unpairing phase transitions.

\begin{figure}[tp]
  \includegraphics[width=0.42\textwidth,clip]{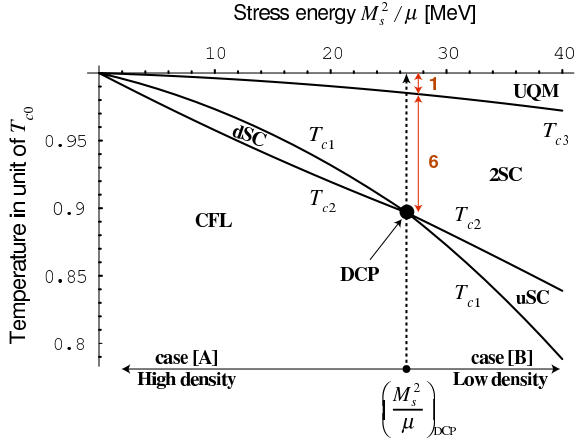}
 \caption[]{%
 The plots of $T_{c\eta}$ evaluated by the Ginzburg-Landau analysis
 with typical parameter set $\De_0=25\MeV$ and $\mu=500\MeV$.
 When the coupling strength is increased,
 the critical DCP stress moves to higher value according to \Eqn{DCPp},
 while the ratio of $T_{c0}-T_{c3}:T_{c0}-T_{\rm DCP}=1:7$ would not be
 affected within the approximation we are working.
 }
 \label{GLschematic}
\end{figure}

Let us now examine the case [B] with a relatively large strange quark
mass.
In this case, $\De_1$ first melts at $T_{c1}=T_{\rm ds}$ and after
that, $\De_2$ disappears at $T_{c2}(>T_{c1})$; that is
\bea
\ba{rcl}
  T_{c1}&=&\dsp T_{c0}\left[1-\log\left(\frac{\mu}{T_{c0}}\right)%
               \frac{\Ms^2}{6\mu^2}-%
               \frac{35\ze(3)\mu^2}{128\pi^2T_{c0}^2}%
               \frac{\Ms^4}{\mu^4}\right],\\[2ex]
  T_{c2}&=&\dsp T_{c0}\left[1-\log\left(\frac{\mu}{T_{c0}}\right)%
               \frac{13\Ms^2}{24\mu^2}-%
               \frac{7\ze(3)\mu^2}{256\pi^2T_{c0}^2}%
               \frac{\Ms^4}{\mu^4}\right].\\[1ex]
\ea
\eea
In this case, the quark matter undergoes another hierarchical
unlocking CFL $\to$ uSC $\to$ 2SC $\to$ UQM.

We have seen that the dSC phase is the second hottest pairing phase
in the case [A], while the uSC phase takes over the dSC phase in the 
case [B].
One can see that the former (latter) situation will be realized at high
(low) density.
In fact, the case [A] includes the high density situation with
the scale hierarchy
\bea
  \frac{\Ms^2}{2\mu}\ll\Ms\ll (T_{c0},\De_0)\ll\mu,
\eea
while the case [B] includes the low density regime with the scale
hierarchy
\bea
  (T_{c0},\De_0),\frac{\Ms^2}{2\mu}\ll\Ms\ll\mu.
\eea 
Thus the case [B] contains an interesting regime with
$\De_{0}\sim{\Ms^2}/{2\mu}$ where the window for the gapless phases open
at $T=0$.
These two physically distinct regimes are separated by the 
{\em doubly critical point} (DCP) which was first discussed
in the numerical analysis of the diquark NJL model
\cite{Fukushima:2004zq}.
We can analytically derive the critical stress for the DCP point
in our current framework as follows;
\bea
  \left.\frac{\Ms^2}{2\mu}\right|_{\rm DCP}(\mu,T_{c0})%
  =\frac{16\pi^2}%
  {21\ze(3)}T_{c0}\left(\frac{T_{c0}}{\mu}\right)%
  \log\left(\frac{\mu}{T_{c0}}\right).\label{eq:DCP}
\eea
Just at the density corresponding to this DCP point,
we have a simultaneous melting of $\De_1$ and $\De_2$, and hence
a direct transition CFL $\to$ 2SC takes place with increasing
$T$ \cite{Fukushima:2004zq}.
We can express critical stress for the DCP in terms of
the CFL gap $\De_0$ at $T=0$ and $\Ms=0$ with the aid of 
the weak coupling universal relation
\bea
  T_{c0}=2^{1/3}\frac{e^{\ga}}{\pi}\De_0%
  \cong0.714\De_0. \label{universal}
\eea
Thus, the DCP point can be parameterized by the 
coupling strength $\De_0$ as
\bea
 \left(\frac{\Ms^2}{2\mu}\right)_{\rm DCP}%
 =\De_0\frac{\De_0}{\mu}%
  \left[1.07+3.19\log\left(\frac{\mu}{\De_0}\right)\right].
\label{DCPp}
\eea
Also, the ratio of $T_{c0}$ to the CFL $\to$ 2SC transition
temperature at the DCP point which we denote by $T_{\rm
DCP}=T_{c1}=T_{c2}$ can be evaluated as
\bea
 \frac{T_{\rm DCP}}{T_{c0}}&=&1-\frac{8\pi^2}%
  {9\ze(3)}\left[\frac{T_{c0}}{\mu}\log\left(\frac{\mu}{T_{c0}}\right)\right]^2\label{eq:TDCP}\\
  &=&1-\left(\frac{\De_0}{\mu}\right)^2\left[0.65+1.93\log\left(\frac{\mu}{\De_0}\right)\right]^2.\nonumber
\eea
We can show the ratio
\beq
  T_{c0}-T_{c3}^{\mbox{\scriptsize[2SC$\to$UQM]}}:T_{c0}-T_{\rm
  DCP}^{\mbox{\scriptsize[CFL$\to$2SC]}}=1:7
\eeq
is universal being independent of the coupling choice $\De_0$.

\begin{table*}[tp]
 \begin{tabular}{|r||c|c|c|c|}
  \hline
  &Regularization scheme & Theoretical parameters  & 
  \multicolumn{2}{c|}{Parameter regime of validity} \\ \hline
  \multicolumn{1}{|c||}{\bf\sl Ginzburg-Landau theory} & mass counter term &
  $T_{c0}, \Ms, \mu$ & $T\sim T_{c0}$,~$\mu\gg\Ms,T_{c0}$ & 
  high density \\ \hline
  \multicolumn{1}{|c||}{\bf\sl Diquark NJL model} & momentum cutoff & 
  $\Lam, G_d, \Ms, \mu$ 
  & $\Lam\agt\mu\agt\mu_{\chi\rm SB}$ 
  & 
  moderate density
  \\ \hline
 \end{tabular}
 \caption[]{
 The comparison of the Ginzburg-Landau 
 theory and the diquark NJL model, \ie, the NJL model without the scalar
 $(q\bar{q})$-coupling
 \cite{Alford:2003fq,Fukushima:2004zq,Fukushima:2005fh}.
 In the derivation of the Ginzburg-Landau potential, we adopt the
 regularization via the mass counter term (Thouless criterion), 
 which insures that all the dimensionful mass scales vanish
 (correlation length diverges) at $T_{c0}$ when $\Ms=0$. 
 This regularization is model-independent in the sense that, after the
 prescription, the effective potential have neither $\Lam$-dependence
 nor the explicit $G_d$-dependence. 
 The Ginzburg-Landau approach is valid for the weak-coupling regime of
 QCD at high density ($\mu\gg\Ms,T_{c0}$), while the diquark NJL
 model is expected to describe the non-perturbavative regime of
 $(\Lam\agt\mu\agt\mu_{\chi\rm SB})$, where $\mu_{\chi\rm SB}$ is the
 chemical potential for the chiral symmetry restoration.
 }
 \label{approaches}
\end{table*}

FIG.~\ref{GLschematic} shows the phase diagram calculated with the
Ginzburg-Landau potential for $\De_0=25\MeV$ and $\mu=500\MeV$,
where the critical temperatures ($T_{c\eta}$) versus the stress energy
$(\Ms^2/\mu)$ is depicted by changing $\Ms$ with $\mu$ fixed.
The DCP is located at $(\Ms^2/\mu)_{\rm DCP}=26.6\MeV\ll\mu=500\MeV$
and $T_c^{\rm DCP}/T_{c0}=0.90$.
These values agree well with those obtained in the numerical analysis 
of the diquark NJL model where $(\Ms^2/\mu)_{\rm DCP}\sim30\MeV$ and
$T_{c}^{\rm DCP}/T_{c0}\sim0.85$ \cite{Fukushima:2004zq}.
If we increase the coupling to $\De_0=(40, 100)\MeV$, then
the DCP shifts to $(\Ms^2/\mu)_{\rm DCP}=(58.5,248.4)\MeV$, and $T_{c}^{\rm
DCP}/T_{c0}=(0.81,0.44)$ according to Eqs.~(\ref{eq:DCP}) and (\ref{eq:TDCP}).
We can see that the agreement with the diquark NJL result
\cite{Fukushima:2004zq} becomes worse in the stronger coupling.
This is because $\Ms^{\rm DCP}$ shifts to higher value as $\De_0$
increases so that the Ginzburg-Landau approach becomes worse due to lack
of the scale hierarchy $\Ms\ll\mu$. 
In TABLE~\ref{approaches}, we have summarized the parameter regime 
where the Ginzburg-Landau approach and the diquark NJL model are valid.
It is worth stressing that according to the Ginzburg-Landau 
evaluation of the DCP given by \Eqn{eq:DCP}, $(\Ms^2/\mu)_{\rm
DCP}$ is located around $\sim \De_{0}(\De_{0}/\mu)$, which is lower with
a factor $\De_{c0}/\mu$ than $\De_{c0}$, \ie, the value corresponding to
the interesting low density regime $\Ms^2/\mu\sim\De_0$; the dSC phase is
realized at higher density than the gCFL/CFL transition density.

\begin{figure*}[thp]
 \begin{minipage}{0.42\textwidth}
  \includegraphics[width=0.9\textwidth,clip]{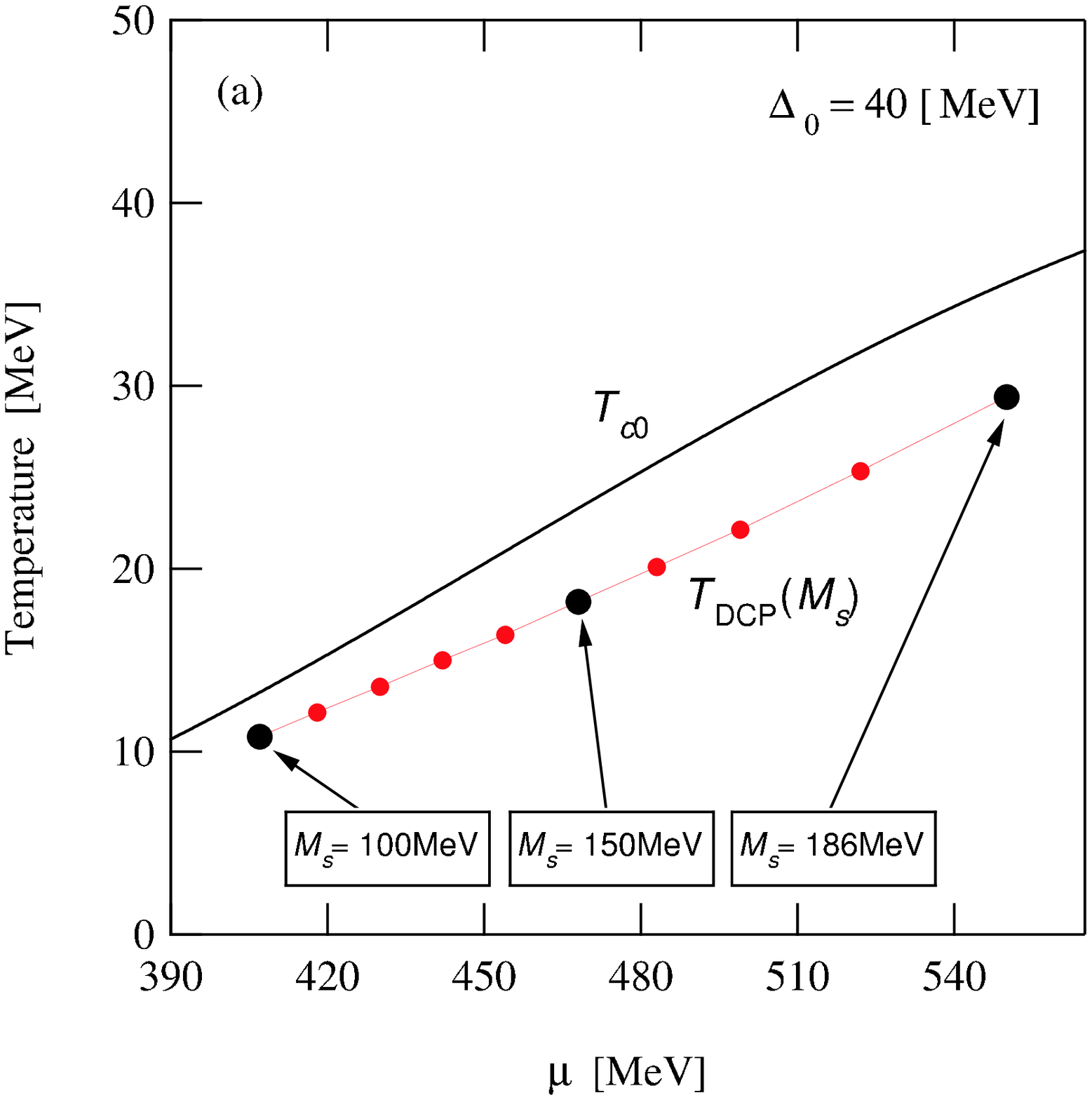}
 \end{minipage}%
 \hfill%
 \begin{minipage}{0.42\textwidth}
  \includegraphics[width=0.9\textwidth,clip]{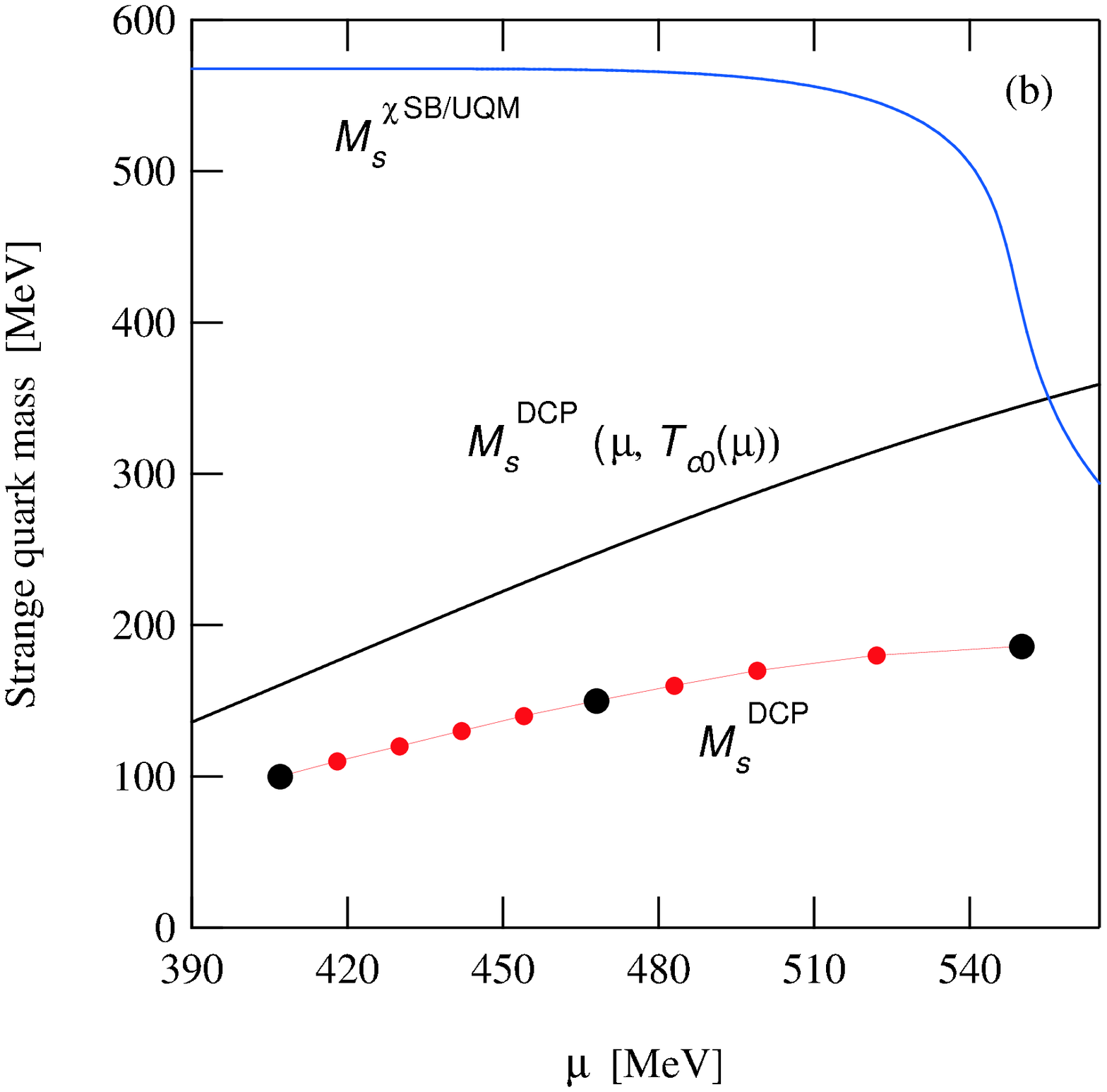}
 \end{minipage}
 \caption[]{
 {\bf (a)}~The DCP point $(\mu,T)_{\rm DCP}$ for various value of 
 $\Ms$; from left to right, $\Ms=100\MeV, 110\MeV, \cdots, 180\MeV$ 
 and $\Ms=186\MeV$. 
 The quark-quark coupling $G_d$ is fixed so that $\De_0=40\MeV$ at
 $\mu=500\MeV$ as in \cite{Fukushima:2005fh}.
 The solid line is $T_{c0}(\mu)$, \ie, the critical temperature for
 $\Ms=0$ as a function of $\mu$.
 {\bf (b)}~The doubly critical strange quark mass $\Ms^{\rm DCP}$ 
 as a function of $\mu$. 
 The solid line indicated by $\Ms^{\rm DCP}(\mu,T_{c0}(\mu))$
 represents the weak coupling Ginzburg-Landau evaluation of $\Ms^{\rm DCP}$ 
 by putting the numerical $\mu$-$T_{c0}$ relation depicted in (a) into
 \Eqn{eq:DCP}. 
 The solid line indicated by $\Ms^{\chi\rm SB/UQM}$ is the dynamically
 determined strange quark mass in the $\chi$SB/UQM sector of our model
 with the scalar $(q\bar{q})$-coupling.
 }
 \label{fig_dsps}
\end{figure*}

In the following, we discuss where in the $(\mu,T)$-plane the DCP is
located.
We first discuss the possibility that it is in the moderate density
regime using the diquark NJL model.
Since the full thermodynamic potential should take the form like
$\Om(\mu,\Ms)=T_{c0}^4 f(\Ms/\mu,T_{c0}/\mu)$ and
$T_{c0}/\mu$ is a slowly varying function of $\mu$,
changing $\Ms$ and changing $\mu$ are independent of each other 
in principle.
If we fix $\Ms$ to some value, we can obtain the phase diagram
in the $(\mu,T)$-plane like FIG.~2 of \cite{Fukushima:2005fh} by
determining the phase by changing $(\mu,T)$; this is contrasted to
FIG.~\ref{GLschematic} where $\Ms$ is changed to control the density.
In this case, we can find the unique DCP in $(\mu,T)$-plane, which
we denote by $(\mu_{\rm DCP},T_{\rm DCP})\equiv(\mu,T)_{\rm DCP}$.
We note that the DCP $(\mu,T)_{\rm DCP}$ shifts in $(\mu,T)$-plane
if $\Ms$ is changed.
We have plotted the DCP $(\mu,T)_{\rm DCP}$ for several choice of
$\Ms$ in FIG.~\ref{fig_dsps}(a).
Although we have used the same diquark NJL model as \cite{Fukushima:2005fh}, 
we did not adopted the {\em chemical potential shift} approximation as
is made there, but performed the exact treatment of $\Ms$ since the
condition $\Ms\ll\mu$ does not hold well.
As can be seen from the figure, as $\Ms$ is increased, the DCP shifts to
larger $\mu$ pushing away the dSC phase to higher density.
We did not find the DCP for $\Ms\agt187\MeV$, where we find only the uSC
in the phase diagram.
This may be a cutoff artifact because $\mu_{\rm DCP}$ increases with
increasing $\Ms$ and approaches $\Lam=800\MeV$.

We have seen that the location of the DCP in the ($\mu,T$)-plane
has one-to-one correspondence to the value of $\Ms$. 
Conversely, if the DCP $(\mu,T)_{\rm DCP}$ is found for some {\em
fixed strange quark mass} $\Ms$, then this $\Ms$ can be viewed as
the doubly critical strange quark mass $\Ms^{\rm DCP}$ for {\em fixed 
chemical potential} $\mu=\mu_{\rm DCP}$;
this follows from the observation that the dSC (uSC) phase will be
realized for $\Ms<\Ms^{\rm DCP}$ ($\Ms>\Ms^{\rm DCP}$) as can be
expected from FIG.~\ref{GLschematic}. 
In this sense, we have plotted the doubly critical strange quark mass
$\Ms^{\rm DCP}$ as a function of $\mu$ in FIG.~\ref{fig_dsps}(b).
For $\Ms$-region above (below) $\Ms^{\rm DCP}(\mu)$ line, the uSC (dSC)
phase would be realized.
Just for comparison, we have also shown the weak coupling
Ginzburg-Landau evaluation of the doubly critical strange quark mass,
which can be obtained via putting $(\mu,T_{c0})$-relation into
\Eqn{eq:DCP} (see solid line indicated by $\Ms^{\rm
DCP}(\mu,T_{c0})$).
We can see the sizable deviation between these two lines, which
is simply attributed to $\Ms\feyn{\ll}\mu$.
However, we stress that the qualitative behaviours are the same; if one
goes higher density, the doubly critical strange quark mass shifts to
larger value so that the dSC phase becomes robuster against the strange
quark mass.
In order to understand why the dSC does not appear in our model with
the scalar coupling $G_s$ and $(q\bar{q})$-condensates, we have also
plotted the dynamically determined strange quark mass in the
$\chi$SB/UQM sector
(see the solid line indicated by $\Ms^{\chi\rm SB/UQM}$). 
The dynamical mass is evaluated on the $\mu$-$T_{c0}$ line in
FIG.~\ref{fig_dsps}(a) so that it should be regarded as the lower
limit of the dynamical strange quark mass in the superconducting phase.
We can see that the line of $\Ms^{\chi\rm SB/UQM}$ is well above the
line for $\Ms^{\rm DCP}(\mu)$ and the two lines never intersect for
$\mu\alt\Lam$, which provides one probable reason for the abscense of
the dSC in the present NJL model with dynamical chiral condensates.

Let us finally discuss the possibility that the DCP is located in 
the weak coupling $(\mu,T)$-regime of QCD.
In QCD, it should be unique because the strange quark mass $\Ms$ is
a decreasing function of $\mu$ approaching its current value $\ms\sim
100\MeV$ in the high density limit, while the doubly critical strange
quark mass $\Ms^{\rm DCP}$ is an increasing function of $\mu$.
If we assume that the crossing (DCP) point $\mu=\mu^*$ at which $\Ms^{\rm
DCP}(\mu^*)=\Ms(\mu^*)$ is located where $\Ms^{\rm
DCP}(\mu^*)\ll\mu^*$, 
then we can estimate $\mu^*$ by the weak coupling Ginzburg-Landau result,
\Eqn{eq:DCP}.
By using the universal relation, \Eqn{universal}, and the weak
coupling perturbative formula for gap, $\De_0(\mu)\sim \mu
g^{-5}e^{-\frac{3\pi^2}{\sqrt{2}g}}$ \cite{Son:1998uk,Schafer:1999jg} as
well, we have the formula $\Ms^{\rm DCP}\sim
\De_{0}(\mu)/\sqrt{\mathstrut g}$; this is actually a slowly increasing
function of $\mu$ when $g$ is identified with the running coupling
constant $\bar{g}(\mu)$ varying with $\mu$.
According to the Schwinger-Dyson analysis of the gap
\cite{Schafer:1999jg,Rajagopal:2000rs,Abuki:2001be},
$\De_{0}(\mu)/\sqrt{\mathstrut \bar{g}(\mu)}$ reaches $\sim100\MeV$
around $\mu\sim10^{10}\MeV$; this indicates $\mu^*$ would be located at
even higher chemical potential since $\ms\sim100\MeV$ is the lower limit
of the mass function $\Ms(\mu)$.
It should be also noted that our starting assumption $\Ms^{\rm
DCP}(\mu^*)\ll\mu^*$ can be justified because the weak coupling
condition $\De_0<\De_0(\mu^*)/\sqrt{\mathstrut\bar{g}(\mu^*)}\ll\mu^*$
is satisfied.
We can conclude that the dSC phase is realized in the extremely high
density regime of $\mu\agt\mu^*$ of QCD.

\vspace*{7ex}
\section{Summary and Outlook}\label{Summary}
In summary, we have investigated the QCD phase diagram with a
special attention to the interplay between the chiral and diquark
dynamics for a wide region of the diquark coupling strength.
Our results for the two limiting cases, the intermediate and strong
coupling, qualitatively agree with those obtained in the recent analyses
\cite{Ruster:2005jc,Blaschke:2005uj}.
Our central results can be summarized as follows.
(i) As the diquark coupling is increased, the phase diagram
gets gradually dominated by the fully gapped phases, while
{\em premature} gapless phases get excluded out as is noted for 
$T=0$ in the previous work \cite{Abuki:2004zk}.
In particular, the phase diagram in the strong diquark coupling
($\De_0\agt160\MeV$) does not include gapless phases and hence
the system is automatically free from the instability problem associated with
imaginary Meissner masses.
(ii) The second hottest pairing phase next to the 2SC phase depends
on the quark density (and also on the diquark coupling strength).
In particular, the dSC is not realized for all the parameter regions 
for which we have performed the calculation.
This can be nicely understood by the Ginzburg-Landau analysis 
incorporating the quartic-order ($\Ms^4$) effects to all the pair
susceptibilities.
In this analysis, we derived the analytical expression for the doubly
critical point (DCP) \cite{Fukushima:2004zq} at which the two
second order critical lines intersect;
one for the 2SC/dSC transition in the high density side and the other
for the 2SC/uSC transition in the low density side.
On the basis of this analysis, we conclude that the window for the
dSC phase at high density $(T_{c0},\De_0)\gg\Ms$ tends to shrink as the
density is lowered.
(iii) We have demonstrated how and why non-trivial first order
transition could be caused by the competition between the diquark
and chiral condensations. 
(iv) We have studied how the 2SC/g2SC and the CFL/gCFL${}_8$ transition
is smeared  by the thermal disturbance by regarding the electron
density and color-neutral isospin density as the order parameters.

In this work, we ignored a Kobayashi-Maskawa-'t Hooft six quark
interaction \cite{Hatsuda:1994pi,Kunihiro:1987bb,Bernard:1987sg}, 
whose effect on the pairing is in part taken into account
in \cite{Ruster:2005jc}.
This term, however, also brings a non-trivial interplay between the
chiral and diquark dynamics through a mixed contraction like,
$|\De_3|^2\Ms$, for example;
these terms indeed will be induced by the procedure of the mean field
approximation.
Also we have neglected a possible $K^0$ condensation in the CFL phase
which may be relevant to the instability problem associated with
the gapless phases \cite{Kryjevski:2004jw,Buballa:2004sx}.
Examining this would require a further improvement of our understanding
on how the meson spectroscopy on the CFL will be modified by the
neutrality constraints \cite{He:2005mp}.
Extending current work so as to take these into account remains indeed
an interesting future problem.

A number of papers have been devoted to the resolution of the
imaginary Meissner masses in the gapless phases
\cite{Unstable,Fukushima:2005cm}. These include;
(i) exploring a possibility of the mixed phase
\cite{Reddy:2004my} along the line suggested earlier
\cite{Neumann:2002jm,Shovkovy:2003ce,Bedaque:2003hi},
(ii) the baryon current generation \cite{Huang:2005pv},
(iii) examining the crystalline pairing phases \cite{Alford:2000ze} 
taking the neutrality constraints into account
\cite{Casalbuoni:2005zp,Giannakis:2005sa}, 
(iv) investigating a possible secondary gap formation
\cite{Hong:2005jv},
(v) the gluon condensation in the g2SC \cite{Gorbar:2005rx},
and 
(vi) an inhomogeneous ($p$-wave) Kaon condensate \cite{Schafer:2005ym}.
Examining all these possibilities or searching for a more 
stable ground state as well as exploring a nature of the 
gluonic instability in the gapless phases is now one of 
central problems in QCD.
Apart from this instability problem, we have shown that the chiral
dynamics in the strange quark sector may simply wash out all the
unstable regime of the gapless phases if the strength of the diquark
coupling is large enough.
It should be noted, however, that the $\mu$-window for g2SC phase still
remains at $T=0$ in the phase diagram even for the diquark
coupling $G_d/G_s=0.5$, \ie, the value extracted from one gluon exchange
vertex. 
Determining the strength of the in-medium diquark coupling from the
phenomenological studies including quark models applied to the exotic
baryon spectroscopy as well as the lattice QCD simulation
\cite{Nakamura:2004ur} will be needed for a realistic description of
the pairing dynamics at low density.

\vspace*{0.3cm}
\noindent
\begin{acknowledgments}
The authors would like to thank M.~Kitazawa for discussions at the 
   early stage of this work.
H. A. is supported by the Fellowship program, Grant-in-Aid for the
   21COE, ``Center for Diversity and Universality in Physics'' 
   at Kyoto University.   
T. K. is supported by Grant-in-Aide for Scientific Research by
   Monbu-Kagaku-sho (No.\ 17540250).
This work is supported in part by a Grant-in-Aid for the 21st Century 
   COE ``Center for Diversity and Universality in Physics''.
\end{acknowledgments}

\vspace*{7ex}
\appendix
\section{Evaluation of Ginzburg-Landau coefficients}\label{GLapproach}
In this section, we present the detail for a derivation of
\Eqn{GLEXP}.
We work with the NJL model for a while, but the results will be found
to be model-independent by a subtraction of the Thouless criterion:
By this subtraction scheme, the model dependence is reduced
only to the critical temperature $T_{c0}$.

We first expand \Eqn{det} in terms of $\De$ at $T\sim T_{c0}$. 
Up to the quartic order in $\De$, we obtain
\bea
 {\mathcal L}_{\rm GL}%
  &=&\Om_e+\frac{4}{G_d}\sum_{\eta=1}^{3}\De_\eta^2\label{det2}\\
  &&-T\sum_n\int\frac{d\bfm{p}}{(2\pi)^3}\tr{\rm Log}\left[%
    {i\om_n\bfm{1}-{\mathcal H}^0_{36}}\right]\nn
  &&+{T}\sum_{m=1}^{4}\frac{1}{m}\sum_{n}%
    \int\frac{d\bfm{p}}{(2\pi)^3}\tr\left[%
    \frac{1}{i\om_n\bfm{1}-{\mathcal H}^0_{36}}%
    {\mathcal V}_{36}\right]^m,\nonumber
\eea
where we have defined
\beq
  {\mathcal H}^0_{36}(\bfm{p};\bfm{M},\bfm{\mu})={\mathcal
  H}_{36}|_{\bfsm{\De}\to0},\,{\mathcal
  V}_{36}(\bfm{\De})={\mathcal H}_{36}-{\mathcal
  H}^0_{36}.\nonumber
\eeq
We omitted the term for the mean-field potential of $\Ms$ in \Eqn{det2}
because we treat it as a constant and as if an external parameter near $T_c$.
The zero-th order term involves the free quark contribution to the
thermodynamic potential and the vacuum fluctuation which we shall simply
subtract.
\beq
\ba{c}
 \dsp %
 -T\sum_n\int\frac{d\bfm{p}}{(2\pi)^3}\tr{\rm Log}\left[%
  {i\om_n\bfm{1}-{\mathcal H}^0_{36}}\right]\\[1ex]
 \dsp %
 =\sum_{a,i}\Om_0(M_i,\mu_{ai},T)%
 -2\sum_{a,i}\int\frac{d\bfm{p}}{(2\pi)^3}E_i,\\[1ex]
\ea
\eeq
where $E_i=\sqrt{p^2+M_i^2}$ and 
\bea
\Om_0(M_i,\mu_{ai},T)&=&-2T\int\frac{d\bfm{p}}{(2\pi)^3}%
 \log\left(1+e^{-(E_i-\mu_{ai})/T}\right)\nn
&&-2T\int\frac{d\bfm{p}}{(2\pi)^3}\log%
\left(1+e^{-(E_i+\mu_{ai})/T}\right).\nonumber
\eea

Because ${\mathcal H}_{36}$ is block-diagonalized as 
${\mathcal H}_{36}={\mathcal H}_{12}^{uds}%
\oplus\sum_{(\al,\be)}^{(db,sg),(sr,ub),(ug,dr)}[({\mathcal
H}^{\al\be}_{4})\oplus(-{\mathcal
H}^{\al\be}_{4})]$, we can
evaluate the quadratic and quartic terms for each sector separately 
as below.

\vspace*{0.3cm}
\paragraph{$({\mathcal H}^{\al\be}_{4})\oplus(-{\mathcal
H}^{\al\be}_{4})$ sector:}
For the symmetry reason, we only have the term which is 
even order in $\De$. 
Thus we have to evaluate the quadratic and quartic terms in \Eqn{det2}. 
The quadratic term can be regarded as the static susceptibility.
For example, we define $\chi^{(db,sg)}$ for the $(\al,\be)=(db,sg)$
sector by
\beq
\ba{c}
\dsp\frac{T}{2}\sum_{n}%
    \int\!\!\!\frac{d\bfm{p}}{(2\pi)^3}\tr\left[%
    \frac{1}{i\om_n\bfm{1}-{\mathcal H}^{\al\be}_{4}}{\mathcal
    V}^{\al\be}_{4}\right]^2
    +(i\om_n\leftrightarrow-i\om_n)\\[4ex]
\dsp=-\De_1^2\chi^{(db,sg)}_{\mbox{\scriptsize
1-loop}}(0,\bfm{0};\mu,T,\mue,\mu_3,\mu_8,\Ms).\\[1ex]
\ea
\eeq
We can analytically evaluate $\chi_{\mbox{\scriptsize
1-loop}}^{(\al,\be)}$ by
expanding it in
$\vec{t}=(\Ms^2/\bar{\mu}_{\al\be},\de\mu_{\al\be})$ with
$\bar{\mu}_{\al\be}\equiv(\mu_\al+\mu_\be)/2$
and $\de\mu_{\al\be}\equiv(\mu_\al-\mu_\be)/2$ being 
referred as the averaged and relative chemical potentials, respectively.
Up to quartic order in $\vec{t}$, we obtain the following
expression by performing all the Matsubara summations and the energy
integrals except for the term suffering from the UV divergence,
\bea
  \chi_{\mbox{\scriptsize 1-loop}}^{(db,sg)}&=&
  2N[\bar{\mu}_{dbsg}]\biggl(1-\frac{\Ms^2}%
  {2\bar{\mu}_{dbsg}^2}\biggl)\int_0\!\!
  dE\frac{1}{E}{\tanh\left(\frac{E}{2T}\right)}\nn
   &&-N[\bar{\mu}_{dbsg}]\frac{7\ze(3)\Ms^4}%
      {32\pi^2\bar{\mu}_{dbsg}^2T_c^2}\nn
   &&-N[\bar{\mu}_{dbsg}]\frac{7\ze(3)%
      \de\mu_{dbsg}\Ms^2}{4\pi^2\bar{\mu}_{dbsg}T_c^2}\nn
   &&-N[\bar{\mu}_{dbsg}]%
     \frac{7\ze(3)\de\mu_{dbsg}^2}{2\pi^2T_c^2}. \label{chi}
\eea
To obtain this, we have ignored the contribution from quasi-antiquark
poles. Also we have defined the density of state by $N[\mu]=\mu^2/2\pi^2$.

The term quartic in $\De_1$ can also be derived within the same
approximation as done in the quadratic term:
\beq
N[\bar{\mu}_{dbsg}]\biggl(1-\frac{3\Ms^2}%
     {4\bar{\mu}_{dbsg}^2}\biggr)\frac{7\ze(3)\De_1^4}{8\pi^2T_c^2}.\label{quartic}
\eeq
We note that order $\vec{t}$ correction corresponding to the
     second term in the above expression gives rise to an order
     $\Ms^4$ correction to the critical temperature $T_{c\eta}$.
However it can be shown that this correction to $T_{c\eta}$
     is suppressed with a factor ${\mathcal O}(T_{c0}^4/\mu^4)$ in
     comparison with that from the $t^2$ terms in the quadratic term in
     \Eqn{chi}.
Therefore we shall ignore the second term in \Eqn{quartic} in the
     discussion below.

We have demonstrated the expansion in the case of $(db,sg)$ sector, but
the Ginzburg-Landau coefficients in other two sectors, $(sr,ub)$ and
$(ug,dr)$, can be derived in the completely same manner.

\vspace*{0.3cm}
\paragraph{${\mathcal H}_{12}^{uds}$ sector:}
Now we take care of the three-flavor mixed sector 
${\mathcal H}_{12}^{uds}$.
The terms quadratic in $\bfm{\De}$ can be shown to have the
completely same form as that in the 
$({\mathcal H}_{4}^{\al\be})%
\oplus(-{\mathcal H}_4^{\al\be})$ sector.
For instance, the term quadratic in $\De_1^2$ becomes
\bea
  -\De_1^2\chi^{(dg,sb)}_{\mbox{\scriptsize 1-loop}},
\eea
where $\chi^{(dg,sb)}_{\mbox{\scriptsize 1-loop}}$ is defined by
\Eqn{chi} with the replacement $(db,sg)\to(dg,sb)$.
On the other hand, the quartic term coming from this sector turns out to
be of the exotic form:
\bea
 &&N[\bar{\mu}_{uds}]\frac{7\ze(3)}%
     {8\pi^2T_c^2}(\De_1^2+\De_2^2+\De_3^2)^2\nn
 &&-N[\bar{\mu}_{uds}]\frac{\Ms^2}{\mu_{uds}^2}\frac{21\ze(3)}%
     {32\pi^2T_c^2}(\De_1^2+\De_2^2)(\De_1^2+\De_2^2+\De_3^2),\nonumber
\eea
where we have defined
$\bar{\mu}_{uds}\equiv(\mu_{ur}+\mu_{dg}+\mu_{sb})/3$.
Also $\Ms^2$ correction can be safely ignored in the analysis below
for the same reason as given previously in the non-mixed two
flavor sector.

\vspace*{0.3cm}
\paragraph{Ginzburg-Landau potential:}
We first investigate the ideal case of $\Ms=\mu_{e,3,8}=0$.
Combining all the results above and putting $\Ms\to 0$ and 
$\mu_{e,3,8}\to0$, we have
\bea
{\mathcal L}_{\rm GL}&=&%
   \sum_{\eta}\De_\eta^2\left[\frac{4}{G_d}%
   -2\chi_{\mbox{\scriptsize 1-loop}}(0,\bfm{0};\mu,T)\right]\nn
  & &+(\De_1^4+\De_2^4+\De_3^4)N[\mu]\frac{7\ze(3)}{8\pi^2T_c^2}\nn
  & &+(\De_1^2+\De_2^2+\De_3^2)^2N[\mu]\frac{7\ze(3)}{8\pi^2T_c^2}.
\eea
In this case all the susceptibilities $\chi_{\mbox{\scriptsize
1-loop}}^{(\al,\be)}$ become of the same form and we denoted them
by $\chi_{\mbox{\scriptsize 1-loop}}$.
The quadratic term is divergent and needs a renormalization.
We renormalize it with the use of the gap equation at $T=T_{c0}$ (The
Thouless criterion for the critical temperature for three massless quark
matter), which plays a role of the mass counter term in \cite{Bailin:1983bm}:
\beq
  \frac{4}{G_d}=2\chi_{\mbox{\scriptsize 1-loop}}%
  (0,\bfm{0},\mu,T_{c0}).\label{subractgap}
\eeq
Subtracting this from the quadratic term, we obtain
\beq
\ba{c}
\dsp\De_\eta^2\left[\frac{4}{G_d}%
     -2\chi_{\mbox{\scriptsize 1-loop}}(0,\bfm{0};\mu,T)\right]\\[3ex]
\to\dsp 4\De_1^2N[\mu]\int_0\!dE\frac{1}{E}%
      \left[\tanh\biggl(\frac{E}{2T_{c0}}\biggr)%
      -\tanh\left(\frac{E}{2T}\right)\right]\\[3ex]
\cong\dsp
        4\De_1^2N[\mu]\log\frac{T}{T_{c0}}%
	\cong4\De_1^2N[\mu]\frac{T-T_{c0}}{T_{c0}}.\\[1ex]
\ea
\eeq
Using this, we obtain
\bea
 {\mathcal L}_{\rm
 GL}&=&(\De_1^2+\De_2^2+\De_3^2)4N[{\mu}]\frac{T-T_{c0}}{T_{c0}}\nn
 &&+(\De_1^4+\De_2^4+\De_3^4)\frac{7\ze(3)N[\mu]}{8\pi^2T_{c0}^2}\nn
 &&+(\De_1^2+\De_2^2+\De_3^2)^2\frac{7\ze(3)N[\mu]}{8\pi^2T_{c0}^2}.
\eea
This is exactly of the same form as that obtained earlier in
\cite{Iida:2000ha,Iida:2002ev}.

We now extend this result with the effects of strange quark mass
and the charge neutrality constraints taken into account.
Our task here is to calculate the splittings of the melting 
($\De_\eta\to 0$) temperature $T_{c\eta}$ up to the quartic order in
$\Ms$.
For this purpose, we need the expansion of the Ginzburg-Landau
coefficients in terms of variables $t=(\Ms^2/\mu,\mue,\mu_3,\mu_8)$.
In order to derive $T_{c\eta}$ up to the quartic order in $\Ms$,
we have to expand the coefficient of the quadratic term up to second
order in $t$.
Thus we start with the following Ginzburg-Landau potential.
\bea
{\mathcal L}_{\rm GL}&=&\Om_e(\mue,T_c)%
     +\sum_{ai}\Om_0(M_i,\mu_{ai},T_c)\nn
  & &+\De_1^2\left(\frac{4}{G_d}%
   -\chi_{\mbox{\scriptsize 1-loop}}^{(dg,sb)}%
   -\chi_{\mbox{\scriptsize 1-loop}}^{(db,sg)}\right)\nn
  & & +\De_2^2\left(\frac{4}{G_d}%
   -\chi_{\mbox{\scriptsize 1-loop}}^{(sb,ur)}%
   -\chi_{\mbox{\scriptsize 1-loop}}^{(sr,ub)}\right)\nn
  & & +\De_3^2\left(\frac{4}{G_d}%
   -\chi_{\mbox{\scriptsize 1-loop}}^{(ur,dg)}%
   -\chi_{\mbox{\scriptsize 1-loop}}^{(ug,dr)}\right)\nn
  & &+(\De_1^4+\De_2^4+\De_3^4)\frac{7\ze(3)N[\mu]}{8\pi^2T_{c0}^2}\nn
  & &+(\De_1^2+\De_2^2+\De_3^2)^2\frac{7\ze(3)N[\mu]}{8\pi^2T_{c0}^2}.  
\eea
$4/G_d$ in the quadratic terms can again be replaced by
$2\chi_{\mbox{\scriptsize 1-loop}}(0,\bfm{0},\mu,T_{c0})$
by using the Thouless criterion in the symmetric matter
\Eqn{subractgap}.
By doing this, the model dependence disappears and only its remnant is
condensed into the {\em parameter} $T_{c0}$. 
After this replacement, we can expand the coefficients of $\De_1^2$,
$\De_2^2$ and $\De_3^2$ in $t=(\Ms^2/\mu,\mue,\mu_3,\mu_8)$.
For example, the $\De_1^2$ term can be expanded up to second order 
in $t$ as follows.
\begin{widetext}
\beq
\ba{rcl}
  \dsp\De_1^2\left(\frac{4}{G_d}%
   -\chi_{\mbox{\scriptsize 1-loop}}^{(dg,sb)}%
   -\chi_{\mbox{\scriptsize 1-loop}}^{(db,sg)}\right)%
  &\to&\dsp
       N[\mu]\De_1^2\biggl[\frac{T-T_{c0}}{T_{c0}}\\[2ex]
  & &\dsp+\left(\frac{2\Ms^2}{\mu^2}%
       -\frac{8\mue}{3\mu}+\frac{2\mu_3}{\mu}+\frac{4\mu_8}{3\mu}\right)\!\!%
  \int_0\!dE\frac{1}{E}\tanh\left(\frac{E}{2T_{c0}}\right)\\[3ex]
  &  &\dsp-\biggl(\frac{3\mu_3+2\mu_8-4\mue}{6\mu}\biggr)^2\!\!%
    \int_0\!dE\frac{1}{E}\tanh\left(\frac{E}{2T_{c0}}\right)%
  \dsp+\frac{7\ze(3)\mu^2}{16\pi^2 T_{c0}^2}%
    \biggl\{\frac{\Ms^4}{\mu^4}%
    +\left(\frac{\mu_3-2\mu_8}{\mu}\right)^2\biggr\}\biggr].\\[2ex]
\ea
\eeq
\end{widetext}
We have used the identity $N[\mu+\de\mu]=%
N[\mu]+\mu\de\mu/\pi^2+\de\mu^2/2\pi^2$.
Due to the asymmetries between cross-species which is caused by $\Ms$
and $\{\mue,\mu_3,\mu_8\}$, we cannot completely eliminate the
divergent quadratic  term by the subtraction of the gap equation at
$T_{c0}$ and we still have divergent corrections proportional to
\beq
  \int_0^{\om_{c}}\!dE\frac{1}{E}\tanh\left(\frac{E}{2T_{c0}}\right)%
 =\log\left(\frac{2\om_c e^\ga}{\pi T_{c0}}\right),
\eeq
with $\ga$ being the Euler constant and $\om_c$ is a UV cutoff for
 the quasi-quark energy.
We note, however, that this divergence is originated simply in our
constant gap parameter ansatz and can be made finite once the momentum
dependence of the gap parameter is properly taken into account; it can
be shown, in the same manner as adopted in \cite{Iida:2003cc},
that we can replace the above divergent contribution by
\beq
  \De_\eta^2\log\left(\frac{2\om_c e^\ga}{\pi
  T_{c0}}\right)%
  \to\frac{1}{2}\De_\eta^2\log\left(\frac{\mu}{T_{c0}}\right)%
  =\De_\eta^2\frac{3\sqrt{2}\pi^2}{4\bar{g}},
\eeq
where $\De_{\eta}$ here should be regarded as the value of the gap
energy at the Fermi-momentum $p=\mu$, and $\bar{g}$ is a running gauge
coupling constant evaluated at energy scale $\mu$. 
In the following, we use the above formula instead of $\om_c$-dependent
expression.
The $\De_2^2$ and $\De_3^2$ sectors can be evaluated in the similar way.
After combining all the results, we have the full Ginzburg-Landau
potential up to the quadratic order in $(\Ms^2/\mu,\mue,\mu_3,\mu_8)$:
\begin{widetext}
\beq
\ba{rcl}
 {\mathcal L}_{\rm
    GL}&=&\Om_e(\mue,T_{c0})+\sum_{ai}\Om_0(M_i,\mu_{ai},T_{c0})\\[2ex]
 &&\dsp+(\De_1^2+\De_2^2+\De_3^2)4N[\mu]\frac{T-T_{c0}}{T_{c0}}%
   +\frac{7\ze(3)N[\mu]}{8\pi^2T_{c0}^2}%
      \left(\De_1^4+\De_2^4+\De_3^4\right)%
    +\frac{7\ze(3)N[\mu]}{8\pi^2T_{c0}^2}%
      \left(\De_1^2+\De_2^2+\De_3^2\right)^2\\[2ex]
 &&\dsp+N[\mu]%
  \biggl[(\De_1^2+\De_2^2)\frac{2\Ms^2}{\mu^2}+%
  \frac{4\mue}{3\mu}\left(-2\De_1^2+\De_2^2+\De_3^2\right)%
  +\frac{2\mu_3}{\mu}(\De_1^2-\De_2^2)%
  +\frac{4\mu_8}{3\mu}(\De_1^2+\De_2^2-2\De_3^2)%
  \biggr]\frac{1}{2}\log\left(\frac{\mu}{T_{c0}}\right)\\[2ex]
  &&\dsp-{N[\mu]}\biggl[\De_1^2\left(%
    \frac{3\mu_3+2\mu_8-4\mue}{6\mu}\right)^2+%
    \De_2^2\left(\frac{3\mu_3-2\mue-2\mu_8}{6\mu}\right)^2%
    +4\De_3^2\left(\frac{\mue-2\mu_8}{6\mu}\right)^2\biggr]%
    \frac{1}{2}\log\left(\frac{\mu}{T_{c0}}\right)\\[2ex]
  &&\dsp+\frac{7\ze(3)N[\mu]}{16\pi^2 T_{c0}^2}\mu^2%
    \biggl[\De_1^2\biggl\{\frac{\Ms^4}{\mu^4}%
    +\left(\frac{\mu_3-2\mu_8}{\mu}\right)^2\biggr\}%
    +\De_2^2\biggl\{\left(\frac{\mu_3+2\mu_8}{\mu}\right)^2+%
    \left(\frac{\Ms^2}{\mu^2}-\frac{2\mue}{\mu}\right)^2\biggr\}%
    +4\De_3^2\left(\frac{\mue^2-\mu_3^2}{\mu^2}\right)\biggr].\\[2ex]
\ea
\label{complicate}
\eeq
\end{widetext}
This is one of the central results from which the analytical
expression for the splittings of the critical temperature ($T_{c0}\to
T_{c\eta}$) can be derived (see Sec.~\ref{GLtheory}).
We now impose the charge neutrality constraints by solving
\bea
  &&\frac{\p{\mathcal L}_{\rm GL}}{\p\mue}=0,%
 \quad\frac{\p{\mathcal L}_{\rm GL}}{\p\mu_3}=0,%
 \quad\frac{\p{\mathcal L}_{\rm GL}}{\p\mu_8}=0,\label{const}
\eea
in $\mue,\mu_3,\mu_8$. 
If $\De_\eta=0$, then the first condition above
can be casted into the following familiar form of the balance equation 
under the weak coupling condition ($T\ll\mu$).
\beq
  N_c\left(\frac{2}{3}\frac{p_{Fu}^3}{3\pi^2}%
  -\frac{1}{3}\frac{p_{Fd}^3}{3\pi^2}%
  -\frac{1}{3}\frac{p_{Fs}^3}{3\pi^2}\right)%
  =\frac{\mue^3}{3\pi^2},\label{balance}
\eeq
with
\beq
\ba{rcl}
  p_{Fu}&=&\mu-\frac{2}{3}\mue,\\[1ex]
  p_{Fd}&=&\mu+\frac{1}{3}\mue,\\[1ex]
  p_{Fs}&=&\sqrt{(\mu+\frac{1}{3}\mue)^2-\Ms^2}.\\[1ex]
\ea
\label{fermimomenta}
\eeq
We can solve \Eqn{balance} order by order in $\Ms$.
Up to the quartic order in $\Ms$, we obtain
\beq
  \mue=\frac{\Ms^2}{4\mu}-\frac{\Ms^4}{48\mu^3}.
\eeq
We here come back to \Eqn{const}, and solve these equations
up to the quadratic order not only in $\Ms^2/\mu$, but also in
$\De_\eta$. We obtain
\bea
  &&\mue=\frac{\Ms^2}{4\mu}-\frac{\Ms^4}{48\mu^3}-\frac{2\De_1^2-\De_2^2-\De_3^2}{6\mu}%
          \log\left(\frac{\mu}{T_{c0}}\right),\nn
  &&\mu_3=\frac{\De_1^2-\De_2^2}{3\mu}%
          \log\left(\frac{\mu}{T_{c0}}\right),\nn
  &&\mu_8=\frac{\De_1^2+\De_2^2-2\De_3^2}{6\mu}%
          \log\left(\frac{\mu}{T_{c0}}\right).\label{chems}
\eea
Substituting these expression to the free potential
$\Om_e+\sum_{ai}\Om_0$ and omitting terms which is independent of
$\De_\eta$, we have the following $\De_\eta^4$ contribution
\beq
\ba{c}
\frac{8\De_1^2(\De_2^2+\De_3^2)-8\De_1^4-5\De_2^4-5\De_3^4+2\De_2^2\De_3^2}%
    {36\pi^2}\log\left(\frac{\mu}{T_{c0}}\right)^2.\\[1ex]
\ea
\nonumber
\eeq
Also the feedback contribution to the quartic term comes from the
quadratic $\De_\eta^2$ terms in \Eqn{complicate}.
Substituting the chemical potentials to the quadratic terms in
\Eqn{complicate} leads to the following quartic terms.
\beq
\ba{c}
  \frac{8\De_1^4+5\De_2^2+5\De_3^4-2\De_2^2\De_3^2%
  -8\De_1^2(\De_2^2+\De_3^2)}{18\pi^2}\log\left(\frac{\mu}{T_{c0}}\right)^2%
  \\[2ex]
  -(\De_2^2-\De_3^2)(-2\De_1^2+\De_2^2+\De_3^2)%
     \frac{7\ze(3)N[\mu]\Ms^2}{48\pi^2\mu^2T_{c0}^2}%
     \log\left(\frac{\mu}{T_{c0}}\right).\\[1ex]
\ea
\nonumber
\eeq
These feedback terms bring about the $\Ms^2$ ($\Ms^4$) corrections to
the critical temperatures $T_{c\eta}$, but these corrections are ${\mathcal
O}(T_{c0}^4/\mu^4)$ suppressed to those from $\Ms^2$ ($\Ms^4$) terms in
the susceptibilities.
Therefore, we can safely ignore these feedback contribution as long as
$\mu\gg T_{c0}$.
Substituting \Eqn{chems} into the quadratic and quartic
terms of \Eqn{complicate} and picking the parts which survives under
$\mu\gg T_{c0}$ lead to our final result for the Ginzburg-Landau
potential, \ie, \Eqn{GLEXP}.
On the basis of this potential, we can argue how the CFL pairing gets
dissolved when $T_{c0}$ is approached as is done in Sec.~\ref{GLtheory}.

In the present framework, we can also give the analytical formula for
the following quantity,
\beq
\ba{rcl}
  \dsp\frac{\De_3^2-\De_2^2}{\De_1^2-\De_2^2}\bigg|_{T\to T_{c0}}&\dsp=&%
  \dsp2+\frac{21\ze(3)}{16\pi^2}%
    \frac{\Ms^2}{T_{c0}^2\log(\mu/T_{c0})}+\cdots.\\[2ex]
\ea
\label{aaa}
\eeq
We have derived this expression starting with the Ginzburg-Landau
potential.
Conversely, if the $\Ms$-dependence of this quantity is known, it
gives some information about the $\Ms$-dependent terms in the
Ginzburg-Landau potential.
In \cite{Fukushima:2004zq}, this quantity is expanded as
$\frac{\De_3^2-\De_2^2}{\De_1^2-\De_2^2}\big|_{T\to
T_{c0}}=a+b\big(\frac{\Ms}{\mu}\big)^2+c\big(\frac{\Ms}{\mu}\big)^4+\cdots$,
and the coefficients, $a,b$ and $c$, are extracted from the numerical
result of the diquark NJL model; $a=2.52$, $b=36.2$ and
$c=1.02\times10^3$ are obtained with a parameter choice
$(\mu,\De_0)=(500,\,25)\MeV$.
We find here, however, that \Eqn{aaa} does not take a form of
a simple expansion in $\big(\frac{\Ms}{\mu}\big)^2$ but rather 
seems to be an expansion in $\frac{\Ms^2}{T_{c0}^2\log(\mu/T_{c0})}%
=\frac{\mu^2}{T_{c0}^2\log(\mu/T_{c0})}\big(\frac{\Ms}{\mu}\big)^2$.
This may explain unusually large numerical values of $b$ and $c$.

\vspace*{7ex}
\section{Off-diagonal color densities}\label{Offdiagonal}
Here, we show that the off-diagonal color densities 
automatically vanish for the standard ansatz with the diquark
condensate,
\ie, \Eqn{model}.
To prove this, we include the chemical potentials not only for the
diagonal but also for off-diagonal color charges as given in
\Eqn{chemicalpot}.
The Nambu-Gor'kov Hamiltonian density in this case takes
the following form after spin-degeneracy removing,
\beq
{\scriptsize
{\mathcal H}_{36}\!=\!\left(%
 \begin{array}{ccccccccc}
 H_{ur} & D_3    & D_2    &        &  M_{12}^\dagger  &        & M_{45}^\dagger &        &      \\
 D_3    & H_{dg} & D_1    & M_{12} &        &        &        &        & M_{67}^\dagger \\
 D_2    & D_1    & H_{sb} &        &        & M_{45} &        & M_{67} &      \\
        & M_{12}^\dagger &        & H_{dr} & -D_3   &        &        &        & M_{45}^\dagger \\
 M_{12} &        &        & -D_3   & H_{ug} &        & M_{67}^{\dagger} &        &      \\
        &        & M_{45}^\dagger  &        &        & H_{sr} & -D_2   & M_{12}^\dagger &      \\
 M_{45} &        &        &        & M_{67} & -D_2   & H_{ub} &        &      \\
        &        & M_{67}^\dagger  &        &  & M_{12}    &        & H_{sg} & -D_1 \\
        & M_{67} &        & M_{45} &        &        &        & -D_1   & H_{db} \\
 \end{array}
\right)
},\label{fullhamiltonian}
\eeq
where all the matrix elements are $4\times 4$ matrices defined by
\beq
\ba{rcl}
H_{ia}\!&=&\!\left(%
\begin{array}{cccc}
  \!M_i\!-\!\mu_{ai}\! & p &   & \\
  p  & \!-M_i\!-\!\mu_{ai} \!&    & \\
     &   & \!M_i\!+\!\mu_{ai}\! & -p \\
     &   & -p & \!-M_i\!+\!\mu_{ai}\!\\
\end{array}%
\right),\\[2ex]
D_\eta\!&=&\!\left(%
\begin{array}{cccc}
    &  &   & -i\De_\eta\\
    &  & -i\De_\eta   & \\
    & i\De_\eta  &  &  \\
 i\De_\eta    &   &  &  \\
\end{array}%
\right),\\[2ex]
M_{\al\be}\!&=&\!\left(%
\begin{array}{cccc}
    \!\!\mu_{\al}\!+\!i\mu_{\be}\!\! &   &   & \\
     & \!\!\mu_{\al}\!+\!i\mu_{\be}\!\!  &   & \\
     &   & \!\!-\mu_{\al}\!+\!i\mu_{\be}\!\!  & \\
     &   &   & \!\!-\mu_{\al}\!+\!i\mu_{\be}\!\! \\
\end{array}%
\right),\\[2ex]
\ea
\eeq
respectively.
We define ${\mathcal H}_{36}^0$ by putting
$\mu_1=\mu_2=\mu_4=\mu_5=\mu_6=\mu_7=0$ in
\Eqn{fullhamiltonian}.
\beq
{\scriptsize
{\mathcal H}_{36}^0\!=\!\left(%
 \begin{array}{ccccccccc}
 H_{ur} & D_3    & D_2    &        &   &        & &        &      \\
 D_3    & H_{dg} & D_1    &  &        &        &        &        & \\
 D_2    & D_1    & H_{sb} &        &        & &        & &      \\
        &  &        & H_{dr} & -D_3   &        &        &        & \\
   &        &        & -D_3   & H_{ug} &        &  &        &      \\
        &        &   &        &        & H_{sr} & -D_2   & &      \\
   &        &        &        &  & -D_2   & H_{ub} &        &      \\
        &        &   &        &  &        &        & H_{sg} & -D_1 \\
        &  &        &  &        &        &        & -D_1   & H_{db} \\
 \end{array}
\right)
},\label{fullhamiltonian0}
\eeq
This Hamiltonian density ${\mathcal H}_{36}^0$ can be further reduced to
the block-diagonalized form as given in the text after some unitary
transformation which makes the Nambu-Gor'kov doubling explicit in the
$(ug,dr)$, $(db,sg)$ and $(sr,ub)$ sectors.
However this is not necessary in the following argument, so
we proceed further with \Eqn{fullhamiltonian0}.

We now define the off-diagonal color charge matrices in the
Nambu-Gor'kov bases as
\beq
  \bfm{Q}_{\al}=-\frac{\p {\mathcal H}_{36}}{\p\mu_{\al}},
\eeq
with $\al=1,2,4,5,6,7$ in addition to the diagonal charges
\beq
  \bfm{Q}_{e,3,8}=-\frac{\p {\mathcal H}_{36}}{\p\mu_{e,3,8}}.
\eeq 

What we have to show is that the off-diagonal color densities
automatically vanish on the ground state determined by ${\mathcal
H}^0_{36}$.
To see this, we first define the complete sets by the following
eigen-value equation:
\beq
  {\mathcal
  H}_{36}^0|p,\al,\si\gt^0=\si\ep_\al^0(p)%
  |p,\al,\si\gt^0.
\eeq
We have 36 eigenvalues which we distinguish by $\al=1,2,\cdots 18$
and the Nambu-Gor'kov spin $\si=\pm$.
The off-diagonal color densities can be written in the form as in
Sec.~\ref{hamil},
\beq
\ba{rcl}
 \rho_\al\!\!&=&\!\!-\frac{1}{2}\sum_{\al=1}^{18}%
             \int\!\!\!\frac{d\bfsm{p}}{(2\pi)^3}%
	     \tanh\left(\frac{\ep_\al^0}{2T}\right)%
	     {}^0\lt p,\al,+|\bfm{Q}_\al%
	     |p,\al,+\gt^0\\[2ex]
             \!\!& &\!\!+\frac{1}{2}\sum_{\al=1}^{18}%
             \int\!\!\!\frac{d\bfsm{p}}{(2\pi)^3}%
	     \tanh\left(\frac{\ep_\al^0}{2T}\right)%
	     {}^0\lt p,\al,-|\bfm{Q}_\al%
	     |p,\al,-\gt^0.
\ea
\label{ocolor}
\eeq
The $\bfm{\Qtilde}$ charge matrix in the Nambu-Gor'kov bases
\beq
\bfm{\Qtilde}=-\bfm{Q}_e-\bfm{Q}_3-\frac{1}{2}\bfm{Q}_8
\eeq
commutes with the Hamiltonian density ${\mathcal H}_{36}^0$, \ie,
$[{\mathcal H}_{36}^0,\bfm{\Qtilde}]=0$, so that the quasi-particle
states $|p,\al,\pm\gt^0$ can be chosen to be the eigenstates of
$\bfm{\Qtilde}$.
Also we note
\beq
\ba{rcl}
  \bfm{Q}_1&=&-i[\bfm{\Qtilde},\,\bfm{Q}_2],\\[2ex]
  \bfm{Q}_2&=&+i[\bfm{\Qtilde},\,\bfm{Q}_1],\\[2ex]
  \bfm{Q}_4&=&-i[\bfm{\Qtilde},\,\bfm{Q}_5],\\[2ex]
  \bfm{Q}_5&=&+i[\bfm{\Qtilde},\,\bfm{Q}_4].\\[2ex]
\ea
\eeq
From these commutation relations, we can immediately conclude 
\beq
\rho_1=\rho_2=\rho_4=\rho_5=0,
\eeq
because
\beq
\ba{c}
  {}^0\lt p,\al,\si|\bfm{Q}_1%
	     |p,\al,\si\gt^0%
  =i\,{}^0\lt p,\al,\si|[\bfm{Q}_2,\,\bfm{\Qtilde}]%
             |p,\al,\si\gt^0=0\\[2ex]
\ea
\eeq
This conclusion follows from $\bfm{\Qtilde}$ is diagonal in the
quasi-particle bases, 
\beq
{}^0\lt p,\al,\si|\bfm{\Qtilde}|p,\be,\si'\gt^0%
=q_\al^\si\de_{\al\be}\de_{\si\si'},
\eeq
where $q_\al^\si$ takes the values $(-1,0,+1)$.

In contrast, the disappearances of $\rho_6$ and $\rho_7$ cannot be
proven in the same manner because of the commutation relations:
\beq
  i[\bfm{\Qtilde},\,\bfm{Q}_6]=0,\quad i[\bfm{\Qtilde},\,\bfm{Q}_7]=0.
\eeq
Thus we try to give a more direct proof here.
We first re-write \Eqn{ocolor} as
\beq
\ba{rcl}
 \rho_\al\!\!&=&\dsp\!\!-\frac{1}{2}\sum_{\si,\al}%
             \int\!\!\!\frac{d\bfm{p}}{(2\pi)^3}%
	     {}^0\lt p,\al,\si|%
	     \tanh\left(\frac{{\mathcal H}_{36}^0}{2T}\right)%
	     \bfm{Q}_\al%
	     |p,\al,\si\gt^0\\[2ex]
            \!\!&=&\dsp\!\!-\frac{1}{2}\sum_{\si,\al}%
             \int\!\!\!\frac{d\bfm{p}}{(2\pi)^3}%
	     {\tr}\left[%
	     \tanh\left(\frac{{\mathcal H}_{36}^0}{2T}\right)%
	     \bfm{Q}_\al\right].\\[1ex]
\ea
\eeq
Since the trace does not depend on the base, we can evaluate
the trace with the {\em natural} base instead of the quasi-particle
eigen-spinors.
Thus what we have to show here turns out to be the proof of
\beq
  {\tr}\left[{\mathcal H}_{36}^0\bfm{Q}_{6,7}\right]=0,\,%
  {\tr}\left[{\mathcal H}_{36}^0{\mathcal H}_{36}^0{\mathcal
  H}_{36}^0\bfm{Q}_{6,7}\right]=0,\,\cdots.
\eeq
If this infinite series of conditions can be shown to be true, 
then we can conclude
\beq
 \rho_6=\rho_7=0.
\eeq

Using the explicit form of the matrices ${\mathcal H}_{36}^0$ and
$\bfm{\Qtilde}_{6,7}$, we have explicitly checked using the {\sl
Mathematica}, that the equation
\beq
  {\tr}\left[({\mathcal H}_{36}^0)^{2n-1}\bfm{Q}_{6,7}\right]=0
\label{conditions}
\eeq
indeed holds for $n=1,2,\cdots,18$.
We have not confirmed \Eqn{conditions} for $n>18$, but the 
confirmation up to $n\le 18$ is adequate for the reason we shall give 
in the following.

First, we write the Hamiltonian density
\beq
\ba{c}
 {\mathcal
 H}_{36}^0=\dsp\sum_{\si,\al}\si\ep_\al(p)P_{\si\al}%
           =\dsp\sum_{\al=1}^{18}(P_{\al}-P_{-\al})\ep_\al(p),
\ea
\eeq
where $P_{\si\al}$ is the projection operator
$|p,\al,\si\gt^0{^0}\lt p,\al,\si|$ in the
natural base; this operator projects vectors out to the eigenspace
in which ${\mathcal H}_{36}^0=\si\ep_\al(p)$ holds.
Using this decomposition, we can re-write the 18 conditions as
\beq
\ba{rrcl}
  1:&0&=&\sum_{\al=1}^{18}\ep_\al%
  \tr\left[(P_\al-P_{-\al})\bfm{Q}_{6,7}\right],\\[2ex]
  2:&0&=&\sum_{\al=1}^{18}\ep_\al^3%
  \tr\left[(P_\al-P_{-\al})\bfm{Q}_{6,7}\right],\\[2ex]
   &\vdots&\\[2ex]
  18:&0&=&\sum_{\al=1}^{18}\ep_\al^{35}%
  \tr\left[(P_\al-P_{-\al})\bfm{Q}_{6,7}\right].\\[2ex]
\ea
\eeq
If all the eighteen roots $\{\ep_\al\}$ with $\al=1,2,\cdots,18$
take different values, then the above eighteen conditions simply mean 
\beq
  \tr\left[(P_\al-P_{-\al})\bfm{Q}_{6,7}\right]=0,
\eeq
for $\al=1,2,\cdots,18$.
Therefore for the arbitrary integer $n$, $\tr[({\mathcal
H}_{32}^0)^{2n-1}\bfm{Q}_{6,7}]=0$ should hold.
In the case that the degeneracy is present as $\ep_1=\ep_2=\ep_3$, for
example, we can prove the following equation in the totally same manner
as in the above argument,
\beq
  \sum_{\al=1,2,3}\tr\left[(P_\al-P_{-\al})\bfm{Q}_{6,7}\right]=0,
\eeq
and again reach the same conclusion $\tr[({\mathcal
H}_{32}^0)^{2n-1}\bfm{Q}_{6,7}]=0$.
Thus, we have proven that the following condition indeed holds for
 $\al=6$ and $7$,
\beq
\ba{c}
\dsp {\tr}\left[\tanh\left(\frac{{\mathcal H}_{36}^0}{2T}\right)%
	     \bfm{Q}_\al\right]=0.\\[1ex]
\ea
\eeq
Consequently, we reached the fact that all the off-diagonal color
densities automatically vanish under the assumption of the diquark
condensate given in \Eqn{model}.

\end{document}